# Remediation of bentazone contaminated water by *Trametes versicolor*: characterization, identification of transformation products, and implementation in a trickle-bed reactor under non-sterile conditions

*Manuel García-Vara [1], Kaidi Hu [2], Cristina Postigo [1*], Lluc Olmo [2], Gloria Caminal [3], Montserrat Sarrà [2], Miren López de Alda [1*]*

*[1]Water, Environmental and Food Chemistry Unit (ENFOCHEM), Department of Environmental Chemistry, Institute of Environmental Assessment and Water Research (IDAEA), Spanish Council for Scientific Research (CSIC), Jordi Girona 18-26, 08034 Barcelona, Spain*

*[2]Departament d'Enginyeria Química, Biològica i Ambiental, Escola d'Enginyeria, Universitat Autònoma de Barcelona, 08193 Bellaterra, Barcelona, Spain*

*[3]Institut de Química Avançada de Catalunya (IQAC), CSIC, Jordi Girona 18-26, 08034 Barcelona, Spain*

***Corresponding authors:***

***Cristina Postigo (cprqam@cid.csic.es), Miren López de Alda (mlaqam@cid.csic.es)***

Institute of Environmental Assessment and Water Research (IDAEA-CSIC)

Department of Environmental Chemistry

C/ Jordi Girona 18-26, 08034 Barcelona, Spain.

Tel: +34-934-006-100, Fax: +34-932-045-904

## ABSTRACT

Bentazone, an herbicide widely applied in rice and cereal crops, is widespread in the aquatic environment. This study evaluated the capacity of *Trametes versicolor* to remove bentazone from water. The fungus was able to completely remove bentazone after three days at Erlenmeyer-scale incubation. Both laccase and cytochrome P450 enzymatic systems were involved in bentazone degradation. A total of 19 transformation products (TPs) were identified to be formed during the process. The reactions involved in their formation included hydroxylations, oxidations, methylations, N-nitrosation, and dimerization. A laccase mediated radical mechanism was proposed for TP formation. In light of the results obtained at the Erlenmeyer scale, a trickle-bed reactor with *T. versicolor* immobilized on pine wood chips was set up to evaluate its stability during bentazone removal under non-sterile conditions. After 30 days of sequencing batch operation, an average bentazone removal of 48 % was obtained, with a considerable contribution of adsorption onto the lignocellulosic support material. Bacterial contamination, which is the bottleneck in the implementation of fungal bioreactors, was successfully addressed by this particular system according to its maintained performance. This research is a pioneering step forward to the implementation of fungal bioremediation on a real scale.

## 1. Introduction

Pesticides are the most predominant chemical substances used in agriculture to prevent and control pests. Although their use has substantial benefits in crop yields and food storage, the environmental damage and human health risks that may result from pesticide application cannot be ignored [1]. Herbicides constitute one of the pesticide groups most frequently used in the EU during the last decade [2]. Among all currently used herbicides, bentazone (3-isopropyl-1H-2,1,3-benzothiadiazine-4(3H)-one-2,2-dioxide) is mainly applied against dicotyledonous weeds in cereals, beans, alfalfa, and other crops worldwide. Different studies have shown that bentazone has a very low rate of mineralization in the environment [3], low sorption to soil (soil adsorption coefficient $K_{oc}$ = 55.3), and relatively high solubility in water (7112 mg $L^{-1}$) [4]. These properties favor leaching (moderate Groundwater Ubiquity Score (GUS) index = 1.95) and run-off of the pesticide from the soil and, thus, explain its ubiquity in groundwater [5–7] and surface water [8–12]. Note that the soil type, pH, and the amount and type of organic material in the soil play also an important role in pesticide mobility, and hence, its fate in the environment. Although bentazone is not highly bioaccumulative, it has also been detected in aquatic organisms [13]. Concerns on the environmental occurrence of bentazone led to consider its inclusion in the list of priority substances in surface water under the EU Water Framework Directive [14].

Under aerobic conditions, bentazone can partially be degraded in soils by microbiota, resulting in the major formation of 8-hydroxy-bentazone, and other transformation products (TPs), namely, 6-hydroxy-bentazone, N-methyl-bentazone, and 2-amino-N-propan-2-ylbenzamide (AIBA) [15]. In the soil upper layers, it is also susceptible to photolysis, transforming into 8-hydroxy-bentazone, among other photoproducts [3]. A considerable higher resistance of bentazone to hydrolytic processes compared with photolytic ones (half-lives of 46-99 days and 2.3-7.5 hours, respectively) has been reported elsewhere [16]. This, together with the aforementioned physical-chemical properties, makes bentazone highly recalcitrant in groundwater [17]. This is confirmed by a numerical model that predicted the presence of bentazone in the aquifer after 20 years since its last field application [18]. Thus, the development of innovative technologies able to reduce or even eliminate bentazone concentrations in the environment is justified and urgent.

Different strategies have been explored for the removal of bentazone from water. Mir *et al.* [19] and Berberidou *et al.* [20] proposed the use of advanced oxidation processes such as $TiO_2$ heterogeneous photocatalysis. The pathways involved in the photocatalytic decomposition of bentazone could be elucidated after identifying 21 phototransformation products formed during the process (e.g., hydroxyl and/or keto derivatives, and dimers) [20]. Compared with chemical and physical treatments, bioremediation appears as a more environmentally friendly and cheaper solution. However, not every bioremediation technique may be suitable for all the contaminants and environmental matrices [21]. For instance, while bentazone persisted in wastewater effluent treated in a pilot-scale membrane bioreactor (MBR) [22], a significant removal was observed in groundwater using different sand filters at full-plant scale [23]. Such a promising result was only observed in one of the three investigated drinking waterworks, which also presented high levels of methane in the groundwater. This finding spurred additional studies that proved the co-metabolic transformation of bentazone into hydroxyl-bentazone by methanotrophic bacteria, in the presence of methane [24]. Biodegradation of bentazone in biological rapid sand filtration occurred mainly through three main biotransformation pathways: oxidation of the isopropyl moiety to the corresponding carboxylic acid, oxidation of the aromatic ring leading to ring cleavage and subsequent decarboxylation, and N-methylation followed by oxidation to a carboxylic acid [25].

White-rot fungi (WRF) are basidiomycetes well-known for their ability to aerobically degrade lignin and various xenobiotics. This is possible thanks to their collaborative group of enzymes including extracellular laccases and peroxidases, and the intracellular cytochrome P450 system [26,27]. Consequently, using the non-specific oxidizing enzymes from WRF seemed to be a smart approach for the elimination of recalcitrant pollutants such as bentazone [28]. In this regard, *Trametes versicolor,* one of the most common WRF species, has been demonstrated to biodegrade a wide number of recalcitrant pollutants including ibuprofen, carbamazepine, atenolol, propranolol, clofibric acid, different estrogens, and even the widespread antimicrobial triclosan [29–32].

The objectives of this work were to determine the ability of the fungus *T. versicolor* to degrade bentazone and to characterize the degradation process in terms of the enzymatic systems responsible for the abatement and the main degradation pathways leading to the identified transformation products (TPs) formed during the process. Furthermore, the stability of a trickle bed reactor (TBR) with immobilized *T. versicolor* to remove bentazone under non-sterile conditions was evaluated.

## 2. Materials and methods

### 2.1. Fungus and culture conditions

*Trametes versicolor* ATCC® 42530™ was acquired from American Type Culture Collection (Manssas, VA) and maintained by subculturing every 30 days on 2% (w/v) malt extract petri dishes (pH 4.5) at 25°C. Blended mycelial suspension and fungal pellets were prepared using malt extract medium (20 g $L^{-1}$, pH 4.5) according to a previously described method [33]. Briefly, four agar plugs of 1 $cm^2$ area from the petri dish grown with the fungi were transferred into 500 mL Erlenmeyer flasks containing 150 mL fresh medium. Then, cultures were incubated at 25 °C under continuous shaking with an orbital shaker (135 rpm). After 5–7 days of incubation, the harvested dense mycelial biomass was blended using an X10/20 homogenizer (Ystral GmbH, Germany), thereby obtaining the mycelial suspension inoculum. Pellets of *T. versicolor* were prepared in 1 L Erlenmeyer flask by inoculating 1 mL of mycelial suspension into 250 mL of fresh medium. After 5–7 days of incubation under shaking condition (135 rpm) at 25 °C, the mycelia pellets were collected and washed with sterile distilled water.

The defined medium used in the Erlenmeyer-scale degradation experiments consisted of 8 g of glucose $L^{-1}$, 3.3 g of ammonium tartrate $L^{-1}$, 1.68 g of dimethyl succinate $L^{-1}$, 10 mL of micronutrients $L^{-1}$, and 100 mL of macronutrients $L^{-1}$ [34]. The pH of the defined medium was adjusted to 4.5 with 1 M HCl or 1 M NaOH. As for TBR experiments, the harvested mycelial suspension was used to inoculate autoclaved pine wood (*Pinus sp.*) chips as reported by Torán et al. [35]. Inoculated pine wood chips were statically incubated for 4 weeks at 25 °C before use.

## 2.2. Chemicals, reagents, and agricultural wastewater

Analytical standards (purity > 99%) of bentazone and its deuterated analog bentazone-$d_7$, dimethyl succinate, commercial laccase purified from *T. versicolor* (20 AU $mg^{-1}$), the laccase mediator 2,2'-azino-bis(3-ethylbenzothiazoline-6-sulphonic acid) diammonium salt (ABTS) (98% pure), and the cytochrome P450 inhibitor 1-aminobenzotriazole (ABT) (98% pure) were purchased from Sigma-Aldrich (Barcelona, Spain). Commercial herbicide KAOS-B (bentazone, 48%) was obtained from SAPEC AGRO (Barcelona, Spain). Formic acid (purity, > 98%) and ammonium acetate were provided by Merck (Darmstadt, Germany). Chromatographic grade acetonitrile (ACN) used for liquid chromatography (LC)-UV analysis was purchased from Carlo Erba Reagents S.A.S. (Val de Reuil Cedex, France). Water and ACN used for ultra-performance liquid chromatography-high resolution mass spectrometry (UPLC-HRMS) analysis were LC-MS grade and supplied by Merck (Darmstadt, Germany) or Thermo Fisher Scientific (USA). A stock solution of bentazone (5 mg $L^{-1}$) was prepared in ethanol and stored at – 20 °C for LC-UV determination of bentazone concentrations. Stock solutions of bentazone and its deuterated analog were prepared in methanol at a concentration of 10 mg $L^{-1}$ for TPs analysis.

The agricultural wastewater used in the TBR was directly collected from an irrigation channel in Gavà agricultural fields, located in the Llobregat River basin (Catalonia, NE Spain), and stored at 4 °C until use. The characteristics of this water are provided as supporting information (SI) in *Table 1*.

**Table 1.** Physicochemical characterization of the agricultural wastewater used in the trickle-bed reactor (TBR).

| Parameter* | Agricultural wastewater** |
|---|---|
| pH | 7.67 ± 0.04 |
| Absorbance at 655 nm | 0.047 ± 0.003 |
| Conductivity (mS $cm^{-1}$) | 2.25 ± 0.07 |
| TSS (mg $L^{-1}$) | 6.33 ± 1.36 |
| VSS (mg $L^{-1}$) | 4.27 ± 1.50 |
| TOC (mg $L^{-1}$) | 16.23 ± 0.81 |
| HPC [lg (CFU $mL^{-1}$)] | 4.68 ± 4.43 |
| Ammonia | n.d. |
| COD (mg $O_2$ $L^{-1}$) | 31.85 ± 0.78 |
| Chloride (mg Cl $L^{-1}$) | 570.50 ± 3.76 |
| Sulfate (mg $SO_4^{-2}$ $L^{-1}$) | 51.24 ± 0.06 |
| Nitrite (mg $NO_2^-$ $L^{-1}$) | 2.78 ± 0.06 |
| Nitrate (mg $NO_3^-$ $L^{-1}$) | 0.08 ± 0.01 |

*TSS: total suspended solids; VSS: volatile suspended solids; TOC: Total organic carbon; HPC: heterotrophic plate counts expressed as the logarithm of colony-forming units per mL, COD: chemical oxygen demand.

** Mean value and standard deviation of triplicate measurements are shown

## 2.3. Degradation experiments in Erlenmeyer flasks

Degradation experiments were performed in 250 mL Erlenmeyer flasks containing 50 mL of fresh defined medium spiked with the commercial bentazone solution at a final concentration of 10 mg $L^{-1}$ of bentazone. This concentration is about 50 times higher than that reported in previous studies in surface waters impacted by rice-growing activities (0.13-0.18 mg $L^{-1}$) [9,36], ¶but it was selected to facilitate the analytical assessment of the system. Briefly, pellets were transferred into each flask as inoculum, achieving a concentration of approximately 2.5 g d.w. $L^{-1}$. Afterward, the cultures were incubated at 25 °C under continuous shaking (135 rpm) for 7 days. Abiotic (non-inoculated) controls, as well as heat-killed culture (121 °C, 30 min) controls, both containing the pesticide were also prepared. All experiments were run in triplicate. Aliquot samples were taken at specific intervals of time during incubation (t = 0, 3, and 7 days) to measure bentazone and glucose concentrations, and laccase activity.

## 2.4. Experiments to evaluate the enzymatic system involved in bentazone degradation

To investigate the role of the different enzymatic systems of *T. versicolor* during bentazone biodegradation, experiments with purified laccase and adding a cytochrome P450 inhibitor were performed

Laccase-mediated degradation experiments were performed in 250 mL Erlenmeyer flasks containing 50 mL dimethyl succinate solution (1.68 g $L^{-1}$, pH 4.5) at a final enzyme activity of 1000 AU $L^{-1}$. The laccase mediator ABTS was added to a final concentration of 0.8 mM [31], to evaluate its effect on bentazone degradation. Abiotic (no laccase) and flasks without the addition of the mediator were also prepared as experimental controls. The flasks were incubated using an orbital shaker (135 rpm) at 25 °C for 24 h. At designated incubation times, 1 mL aliquots were collected and mixed with 100 µL of 1 M HCl to stop the reaction to measure bentazone concentration. After 24 h, 1 mL of the culture was withdrawn from the laccase control to measure the enzymatic activity.

Experiments designed to evaluate the effect of inhibiting the cytochrome P450 system on bentazone degradation were also performed in Erlenmeyer flasks containing 2.5 g d.w. $L^{-1}$ of fungal pellets. 1-aminobenzotriazole was added as an inhibitor, to a final concentration of 5 mM [31]. Then, pellet cultures were incubated for 42 h at 25 °C under continuous shaking (135 rpm). An experimental control that consisted of the pellet culture incubated in the absence of an inhibitor was run in parallel. Each experimental condition was conducted in triplicate. Aliquot samples were collected at designated times during the culture incubation for the analysis of bentazone concentration.

**2.5. Degradation experiments for transformation products identification**

Biodegradation experiments for the identification of TPs were carried out in 500 mL Erlenmeyer flasks, with 100 mL of fresh defined medium spiked with analytical-grade bentazone at a final concentration of 1 mg $L^{-1}$, and a concentration of fungal pellets of 2.5 g d.w. $L^{-1}$. Cultures were incubated in the dark at 25 °C under agitation (135 rpm) for 7 days. Abiotic (non-inoculated) and heat-killed culture (121 °C during 30 min) solutions, both containing 1 mg $L^{-1}$ of bentazone were used as experimental controls. A solution containing fungal pellets but no bentazone was used also as a control to detect potential artifacts formed during fungal degradation. 4 mL samples were taken at 0 h, 6 h, 11 h, 24 h, 3 d, and 7 d, centrifuged at room temperature at 17,700 × g for 4 min, and stored at -20 °C until UPLC-HRMS analysis. All experimental conditions were run in triplicate.

## 2.6. Degradation experiments in a TBR bioreactor

A cylindrical TBR filled with pre-inoculated pine wood chips was set up (*Figure 1*). One liter of agricultural wastewater (Table 1) fortified with the commercial bentazone solution to a final concentration of 10 mg $L^{-1}$ of bentazone was loaded into the packing bed at the top of the reactor through a rotary distributor, and then collected in the reservoir tank placed at the bottom. The tank was equipped with a magnetic stirrer and a pH controller, which assisted in keeping the pH of the solution at 4.5 by adding 1 M HCl or 1 M NaOH during the duration of the experiment. To assess the role that adsorption by the lignocellulosic support plays, an identical reactor filled with non-inoculated pine wood chips was set as a control. Multiple runs were operated in sequencing batch reactor (SBR) mode with a 3-days cycle at room temperature. Wastewater was renovated after each batch experiment and a recycling ratio (RR) of 300 was adopted. Samples were taken from the tank at designated time intervals to measure bentazone concentration, laccase activity, heterotrophic plate counts (HPC), and chemical oxygen demand (COD).

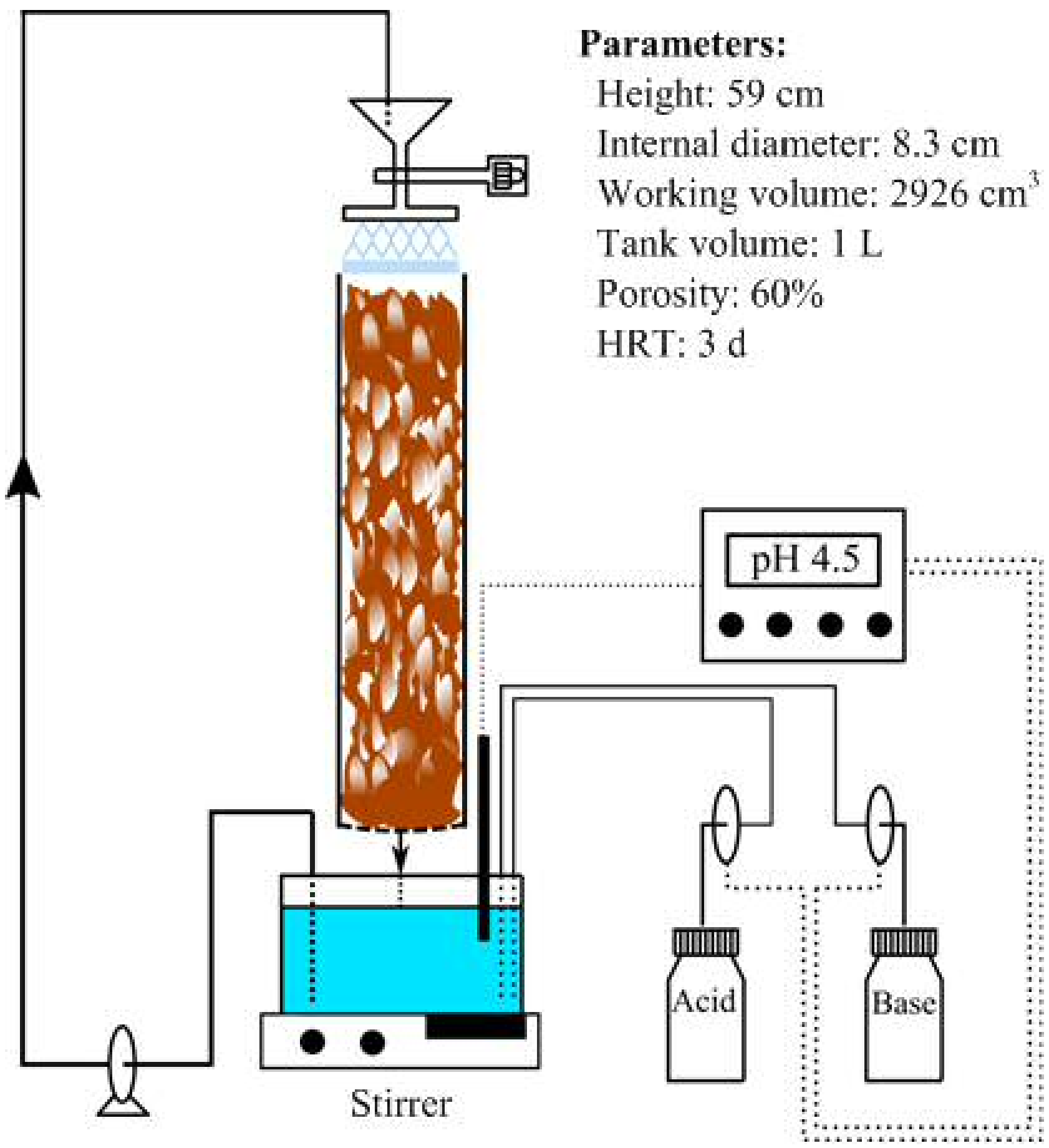

**Figure 1**. Schematic representation of the experimental TBR set-up. HRT, hydraulic retention time.

## 2.7. Analytical methods

### *2.7.1. Laccase activity*

Laccase activity was measured through the oxidation of 2,6-dymetoxyphenol (DMP) by the enzyme in the absence of a cofactor, using a modified version of the method for the manganese peroxidase system [37]. Activity units per liter (AU $L^{-1}$) are defined as the amount of DMP in µM oxidized per minute. The molar extinction coefficient of DMP was 24.8 $mM^{-1}$ $cm^{-1}$.

### *2.7.2. Bentazone concentration*

Samples were firstly filtered through a Millipore Millex-GV PVDF membrane (0.22 µm). Then the residual bentazone concentration was determined using an HPLC Ultimate 3000 (Dionex, USA) equipped with a UV detector. Chromatographic analysis was achieved with a C18 reversed-phase column (Phenomenex®, Kinetex® EVO C18 100 Å, 4.6 mm × 150 mm, 5 µm), kept at 30 ºC, and a mobile phase consisting of water, containing 0.01% formic acid, (v/v) (A) and acetonitrile (B) at a constant flow rate of 0.8 mL $min^{-1}$. The organic gradient for chromatographic separation was as follows: 35% of B from 0 min to 5 min, then a linear increase of B to 45% from 5.01 min to 15 min, return to initial conditions in 1 min, and maintenance of initial conditions for 2 more min. The injection volume was 40 µL. The detection wavelength was set at 254 nm. The limit of detection was 0.5 mg $L^{-1}$.

### *2.7.3. Analyses for agricultural wastewater characterization*

The absorbance at 650 nm was determined by a UNICAM 8625 UV/VIS spectrometer, and the conductivity was monitored by a CRISON MicroCM 2100 conductometer. The total suspended solids (TSS) and volatile suspended solids (VSS) were measured according to the standard methods 2540 D and 2540 E, respectively [38]. The total organic carbon (TOC) was determined using an Analytik Jena multi N/C 2100S/1 analyzer. The HPC results were reported as the logarithm of colony-forming units (CFU) per mL [lg (CFU $mL^{-1}$)] using the spread-plate method with a plate count agar (PCA) following the standard method 9215 [38]. The N-$NH_4^+$ concentration and the COD were analyzed using commercial kits LCK 303 and LCK 314 or LCK 114, respectively (Hach Lange, Germany). Chloride, sulfate, nitrite, and nitrate anions were measured by ion chromatography using a Dionex ICS-2000 equipped with Dionex IonPac AS18-HC column (250 mm x 4 mm) which was eluted at 1 mL min$^{-1}$ with a 13 mM KOH aqueous solution.

### *2.7.4. Identification of TPs*

The analysis of the biotransformation products formed during bentazone degradation was done with a UHPLC system Acquity (Waters, Milford, MA, USA) coupled to a hybrid quadrupole-Orbitrap Q-Exactive mass spectrometer (Thermo Fisher Scientific, San Jose, CA, USA). Sample components were separated on a chromatographic column Purospher® STAR RP-18 endcapped Hibar® (150 x 2.1mm, 2µm) (Merck, Darmstadt, Germany), using an organic gradient of a mobile phase consisting of water with 20 mM of ammonium acetate (A) and acetonitrile (B) at a constant flow rate of 0.3 mL min$^{-1}$. After 1 min of isocratic conditions (5% of acetonitrile), the proportion of the organic component increased linearly to 20% in 2 min, to 80% in the next 3 min, and 100% in one more min. After 2 minutes of static conditions, a fast gradient restored the mobile phase to initial conditions (in 0.5 min), which were maintained during 4 min for column re-equilibration. The injection volume was set to 10 µL.

The HRMS analysis was performed using a heated electrospray ionization (HESI) source operated in the negative ion mode. Ion source conditions were: capillary voltage, -2500V; temperature, 350°C; sheath gas flow rate, 40 arbitrary units; auxiliary gas, 10 arbitrary units; vaporizer temperature, 400°C. Nitrogen (>99.98%) was employed as the sheath, auxiliary, and sweep gas. Accurate mass measurements were done in data-dependent acquisition mode. First, a full MS scan was conducted using a full width at half maximum (FWHM) resolution of 70,000 (at *m/z* 200). The *m/z* range covered expanded from 70 to 600 to include also phase II metabolites. Then, data-dependent MS/MS scan events were recorded for the n=5 most intense ions (>$10^5$ counts) detected in each scan with an FWHM resolution of 17,500 (at *m/z* 200) and using a normalized collision-induced dissociation energy of 40. MS data acquisition was done using Xcalibur v4.1.

Data processing was conducted with Compound Discoverer v.3.1 (Thermo Fisher Scientific, San Jose, CA, USA). This software was used for peak alignment and deconvolution (with a retention time maximum shift of 2 minutes and a mass tolerance of 5 ppm), feature grouping, and elemental composition prediction. A handmade MS library of suspect compounds, that included bentazone TPs published in the peer-reviewed literature [3,15,16,20] and those obtained with the EAWAG-BBD Pathway Prediction tool [39], and different chemical compound databases (ChemSpider, mzCloud, mzVault) were used to assign potential compound identities. Then, the *m/z* list initially generated was manually revised to search for potential TP candidates (i.e., those present only in experimental reactors at t = 3 and 7 days and absent in control and blank samples), and their $MS^2$ spectra were examined for structure identification. Fragment rationalization and structure proposal of the TPs identified were supported by the software ChemDraw Professional v18.1 (PerkinElmer Informatics).

### *2.7.5. Data analysis*

The degradation efficiency of the different investigated systems was evaluated through the percentage of bentazone remaining in the solution, according to the equation:

$$\text{Degradation percentage} = \frac{C_0 \quad C_t}{C_0} \times 100\%$$

where $C_0$ is the initial bentazone concentration at $t_0$ and $C_t$ corresponds to the residual bentazone concentration in the culture at a given time t. The mean and standard deviation (SD) of triplicate measurements were calculated. The statistical significance of the changes observed, with a level of confidence of 95% ($\alpha$=0.05), was determined using SPSS v22.0.

## 3. Results

### 3.1. Degradation of bentazone by *T. versicolor*

The capability of *T. versicolor* to degrade bentazone was firstly evaluated under sterile conditions at Erlenmeyer-scale. *T. versicolor* completely removed 10 mg $L^{-1}$ of bentazone within 3 days (*Figure 2*). Comparing bentazone concentration decay in the abiotic and the heat-killed controls, 9% of the bentazone removal could be ascribed to adsorption of the compound onto the biomass. A maximum laccase activity (9.12 AU $L^{-1}$) was also achieved after 3 days and then reduced during incubation. Concerning glucose, it was almost completely utilized after 3 days.

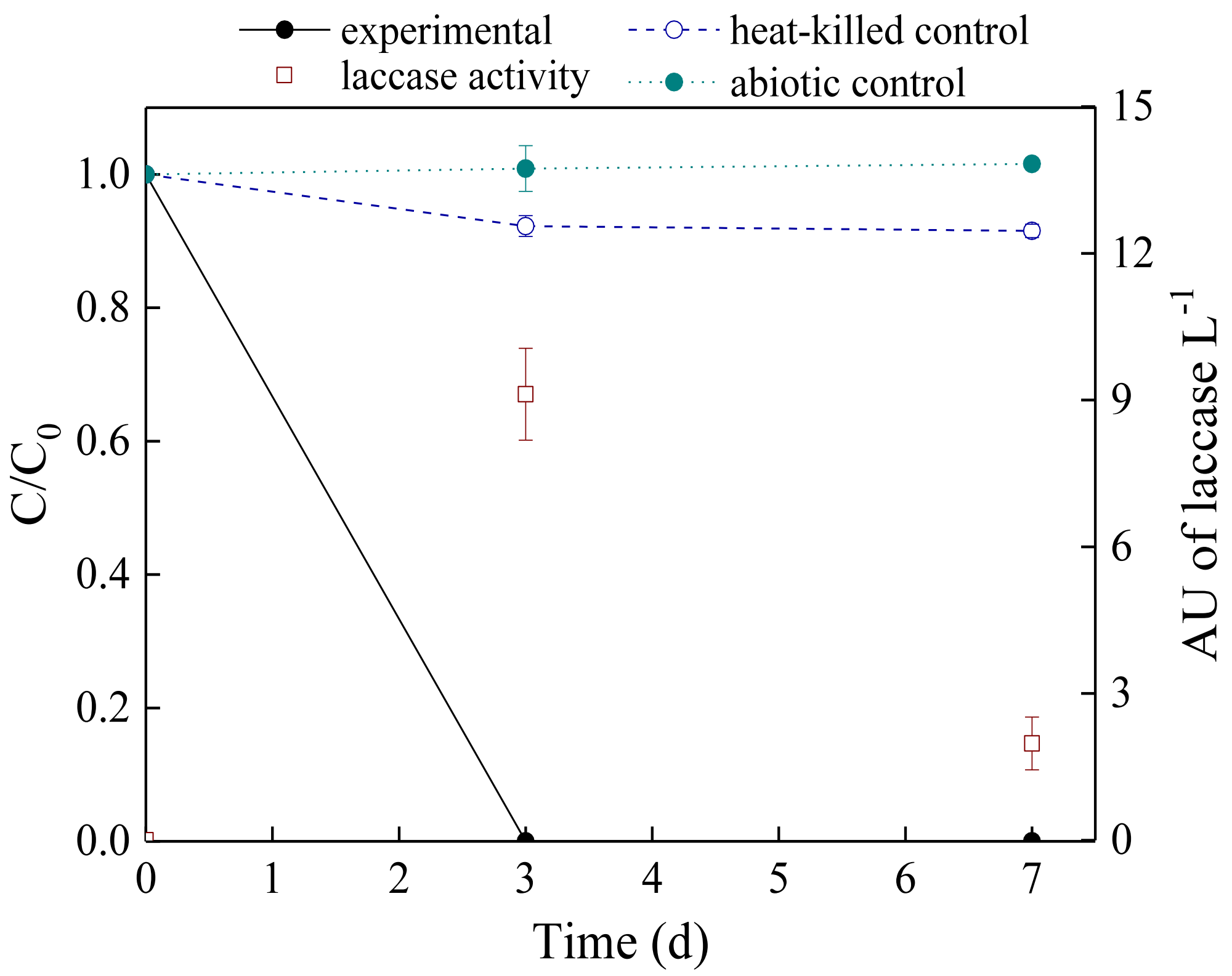

**Figure 2.** Time-course of bentazone degradation by *T. versicolor* (y-axis on the left) and laccase activity during the process (y-axis on the right). **C** represents the residual concentration of bentazone in the sample (mg $L^{-1}$), and $\mathbf{C_0}$ corresponds to the concentration of bentazone at the beginning of the experiment (10 mg $L^{-1}$). Values are the mean and standard deviation of three replicates.

### 3.2. The role of laccase and cytochrome P450 enzymatic systems in the degradation of bentazone

Both laccase and cytochrome P450 systems have been reported to be involved in the degradation of recalcitrant environmental pollutants by filamentous fungi [26,40]. Laccase, an extracellular enzyme, is a good indicator of fungal activity during the degradation process. The cytochrome P450 system, an intracellular enzymatic system, has been reported to catalyze the first detoxification step towards a wide range of toxic compounds in mammals and therefore, it could also play a key role in the degradation process of bentazone.

The participation of the laccase enzymatic system in bentazone degradation was investigated with *in vitro* experiments using commercial laccase. Results showed that the laccase system could degrade bentazone completely in the presence of the ABTS mediator after 1 hour of treatment (*Table 2*). On the contrary, when ABTS was not present in the solution, only 11% of the initial bentazone concentration was degraded in the same period and this figure did not improve in 24 h of treatment. However, laccase was still active at the end of the incubation period (100.38 AU $L^{-1}$). Thus, our findings indicate that laccase is involved in the biotransformation process.

**Table 2.** Bentazone degradation by laccase enzymatic system in the presence and absence of ABTS.

| Time | $C/C_0$ | | |
|---|---|---|---|
| | Abiotic control | Laccase | ABTS |
| 0 h | 1 | 1 | 1 |
| 30 min | 0.97 ± 0.039 | 0.88 ± 0.039 | 0.068 ± 0.0079 |
| 1 h | 0.96 ± 0.037 | 0.89 ± 0.037 | 0 |
| 2 h | 0.98 ± 0.041 | 0.89 ± 0.038 | 0 |
| 4 h | 0.98 ± 0.039 | 0.88 ± 0.039 | 0 |
| 6 h | 0.98 ± 0.041 | 0.91 ± 0.013 | 0 |
| 10 h | 0.97 ± 0.042 | 0.89 ± 0.037 | 0 |
| 24 h | 0.98 ± 0.041 | 0.88 ± 0.024 | 0 |

Note: Each value represents the mean of triplicate measurements ± SD.

To investigate the role of the cytochrome P450 enzymatic system in bentazone degradation, *in vivo* experiments were conducted in the presence of the cytochrome P450 inhibitor, 1-aminobenzotriazole. Bentazone degradation was considerably inhibited when 1-aminobenzotriazole was added into the system (*Figure 3*). The initial decrease in bentazone concentration observed at the beginning of the experiments could be attributed to adsorption onto the fungal pellets (in line with the results obtained in the initial degradation experiments, section 3.1). In any case, it is clear that cytochrome also participates in bentazone degradation by *T. versicolor*.

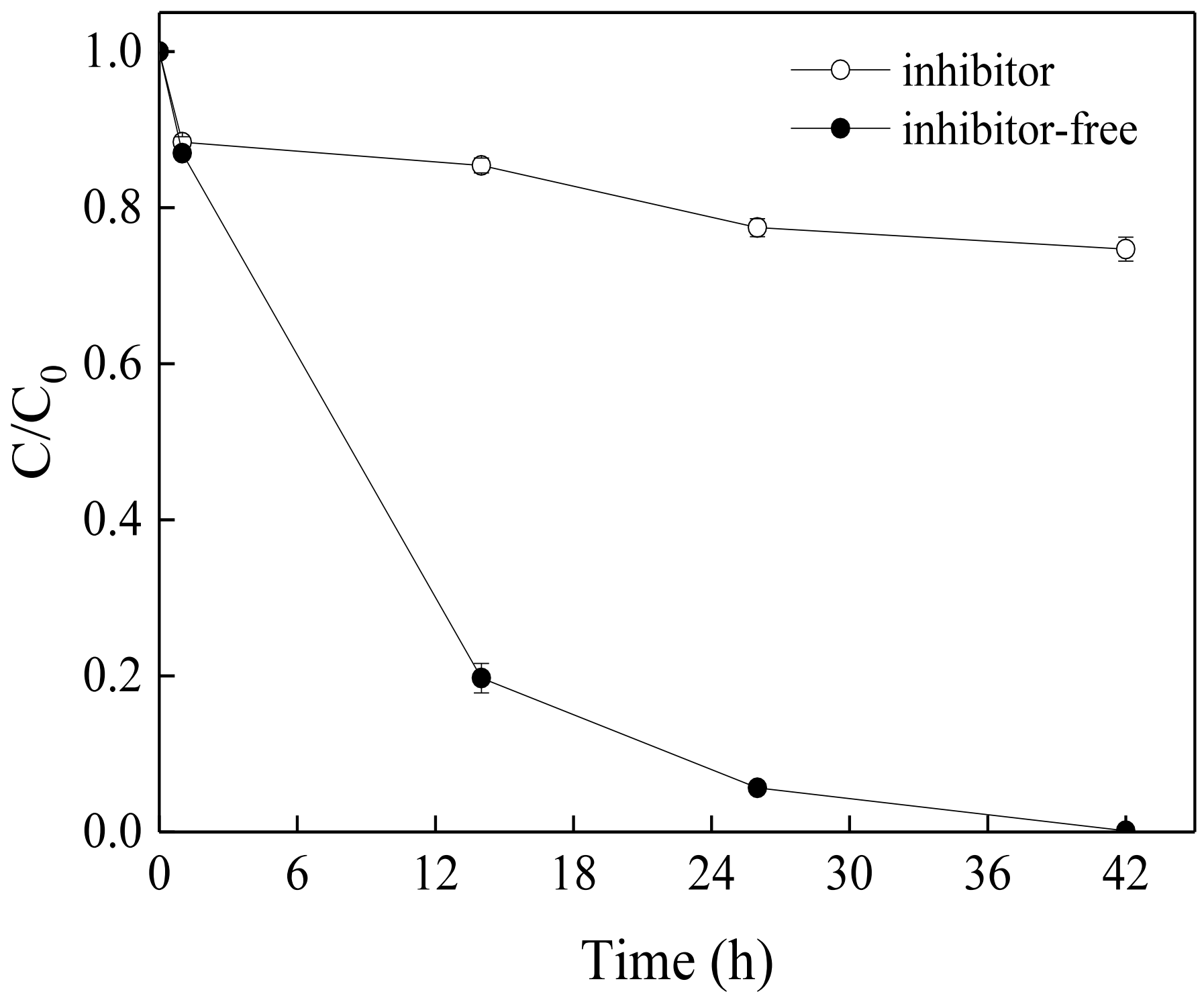

**Figure 3**. Influence of the cytochrome P450 inhibitor 1-aminobenzotriazole on the degradation of bentazone by *T. versicolor*. **C** represents the residual concentration of bentazone in the sample (mg $L^{-1}$), and **$C_0$** corresponds to the concentration of bentazone at the beginning of the experiment (mg $L^{-1}$). Values are the mean and standard deviation of three replicates.

### 3.3. Identification of TPs and main degradation pathways

The identification of TP candidates in Compound Discoverer-processed HRMS data required a filtration step. For this, different criteria were used including i) the ratio of the *m/z* feature between time zero and the different sampling times, ii) the presence/absence of the molecular ions in the abiotic and heat-killed controls, and iii) the retention time.

Main TPs and corresponding characteristic fragment ions detected using the workflow described in section 2.7.4. and the aforementioned filtration step are shown in *Table 3.* Based on the TPs identified, a biodegradation pathway has been proposed (*Figure 4)*. In total, 19 molecular ions were identified as TPs by UPLC-HRMS; however, a chemical structure could be only tentatively proposed for 8 of them. Their corresponding full MS chromatograms and $MS^2$ spectra are provided in *Figures S1-S19* as SI, and their time-evolution is shown in *Figure 6* As chemical structures are not confirmed with the analysis of reference standards, they are proposed with a confidence level of 3 in all cases except for TP494, for which no $MS^2$ data were obtained and, therefore, fits a confidence level of 5 [41]. As shown in Figure S2 in SI, bentazone ($t_R$ = 5.10 min) gets deprotonated under HESI(-) conditions (*m/z* 239.0496). Upon collision-induced dissociation of the precursor ion, fragment ions were detected at *m/z* 197.0021, *m/z* 175.0877, and *m/z* 132.0329, which correspond to the loss of the isopropyl moiety, the loss of the sulfonyl group, and the combined loss of the aforementioned moieties, respectively.

Although N-methyl-bentazone, a bentazone TP commonly found in the environment [3,15], was not detected during the degradation experiments, TP268 and TP284a were identified as oxidized forms of N-methyl-bentazone. TP268 appeared after 5 h while TP284a was formed at a later stage, which could indicate that TP284a results from the oxidation of TP268 (*Figure 4).* This kind of tertiary carbon oxidation is a reaction frequently attributed to the cytochrome P450 system [42]. Moreover, cytochrome P450 may also catalyze the hydroxylation of the aromatic ring, resulting in the formation of TP256 (hydroxyl-bentazone, $t_R$ = 4.62 min). 6-OH and 8-OH-bentazone are typical TPs formed by the soil microbiota metabolism [3,15]; however, with the available structural information, it is not possible to indicate the isomer that corresponds to the TP formed during fungal degradation. Similar to TP284a, TP284b is also proposed to be formed after the carboxylation of bentazone. Due to its fragmentation pattern, carboxylation of TP284b could only be possible at the isopropyl moiety (*Figure 4)*. The transformation of bentazone into TP285 was tentatively produced by hydroxylation and N-nitrosation of the free secondary amine of the parent compound. Microbiologically induced N-nitrosation has been reported for different species [43,44] and resulted in the formation of stable products after secondary amine nitrosation [45]. Although this TP has not been confirmed yet, it is of special concern, because most N-nitroso compounds are classified as carcinogens [46]. Main transformation pathways, including oxidation of the isopropyl moiety, N-methylated oxidations, or hydroxylations of the aromatic ring are in line with those reported during microbial biodegradation in sand filters [25].

The laccase catalytic mechanism consists of the abstraction of an electron from a substrate to produce free radicals [27,47]. TP258 and TP286 seem to be formed from a hydroxylated form of bentazone through this laccase-mediated mechanism (*Figure 4*). The formation of a radical in one of the hydroxyl groups of the aromatic ring could derive in the electronic rearrangement to produce an iminoquinone intermediate. Then, a nucleophilic attack of a free $HO^•$ and the subsequent intramolecular ring closure results in the formation of TP258 (*Figure 5*). Later, the addition of the hydroxyl-methyl group that had previously been eliminated from the iminoquinone intermediate could generate TP286. Regarding TP494, no $MS^2$ data were obtained; however, laccase induced dimerization of bentazone could be a plausible reaction already observed in nature [15]. Most of the remaining TPs were grouped on phase II metabolites, although no logical structure could be proposed for them. Their fragmentation pattern showed a common fragment ion at *m/z* 239.0496, corresponding to the molecular ion of bentazone. HRMS data provided a good insight into the biodegradation pathway carried out by *T. versicolor*; however, further work is needed for the identification of those TPs for which a plausible structure could not be proposed.

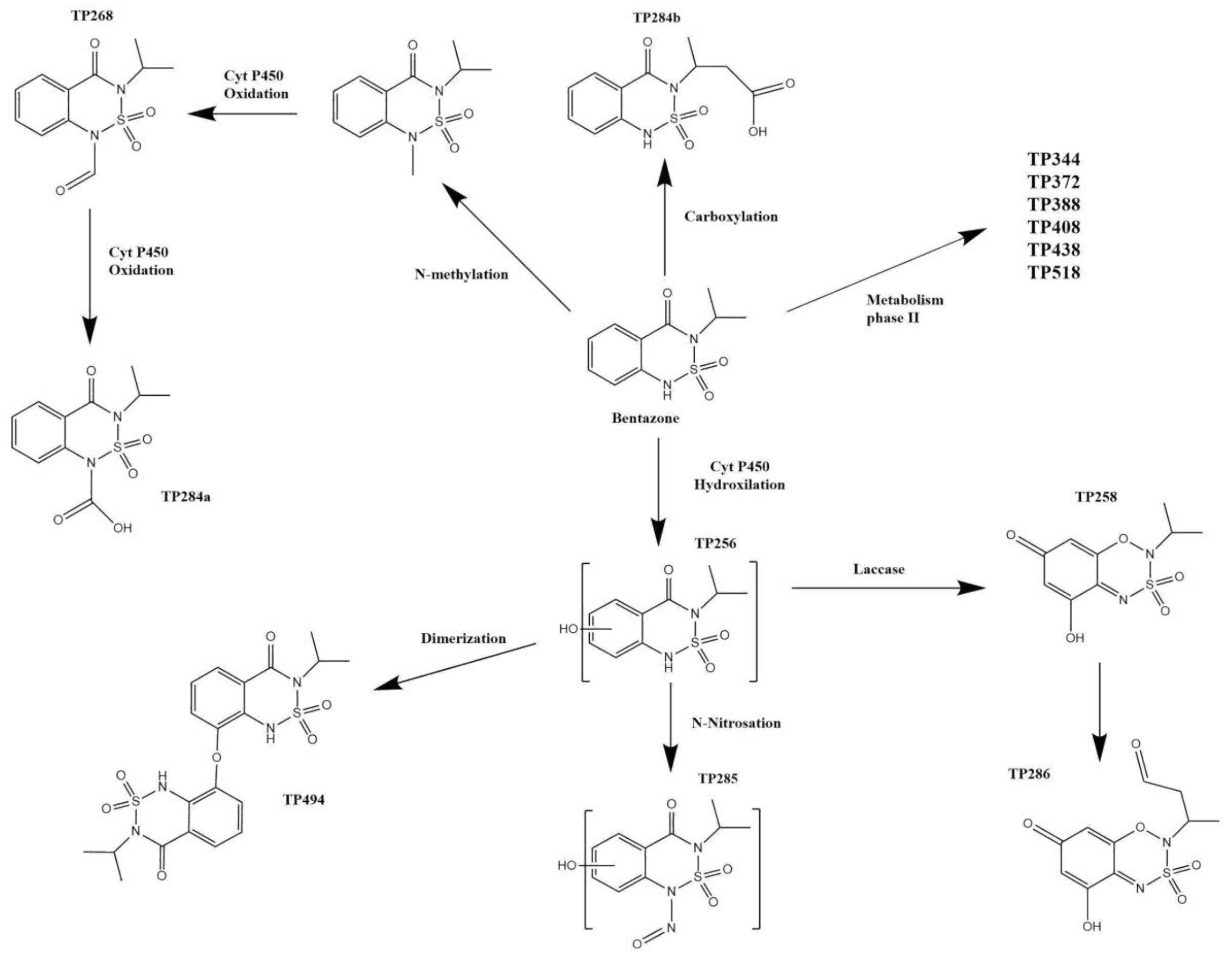

**Figure 4.** Bentazone biodegradation by *T. versicolor*

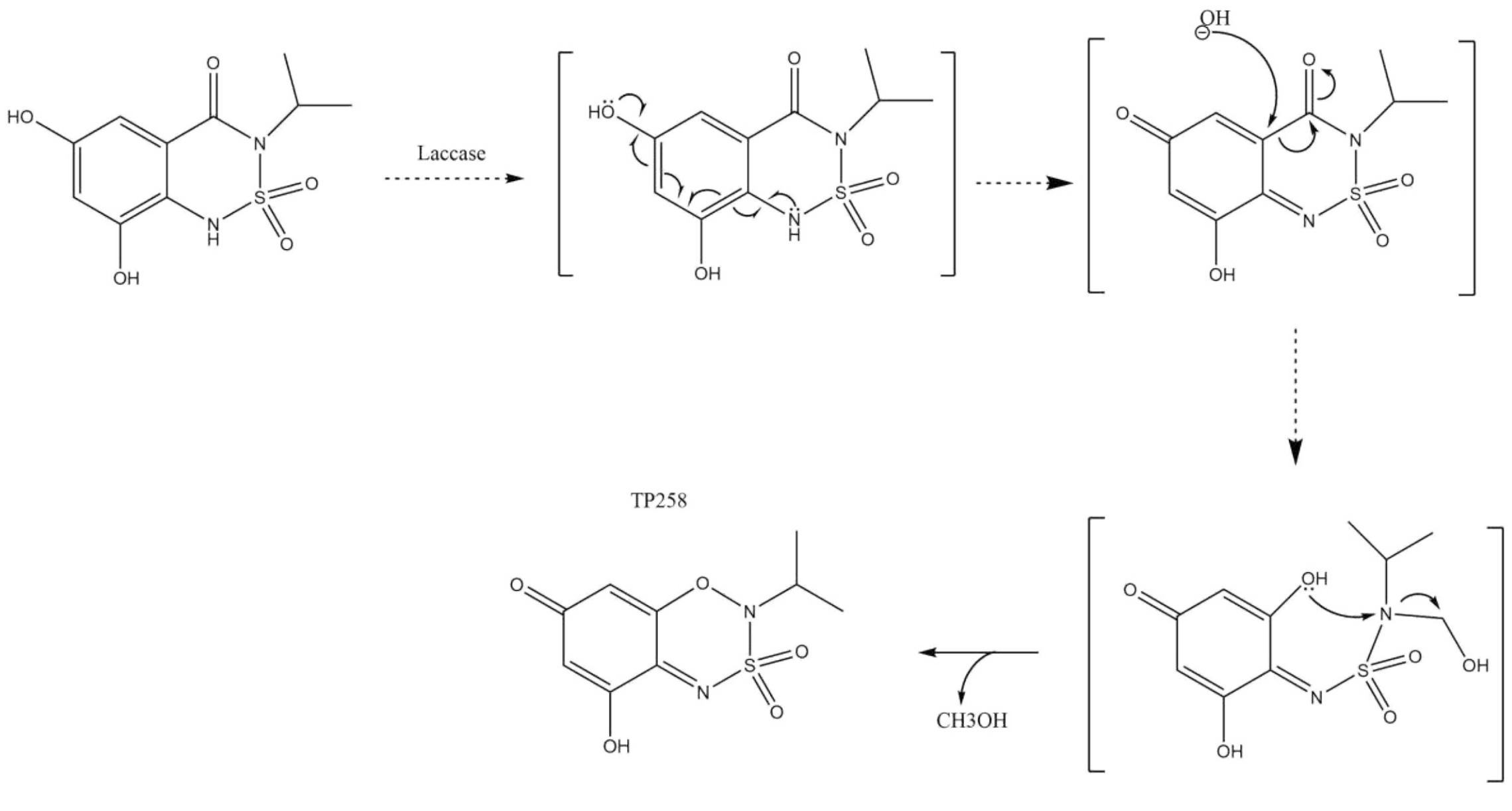

**Figure 5.** Proposal of the laccase mediated radical reaction involved in the formation of TP258.

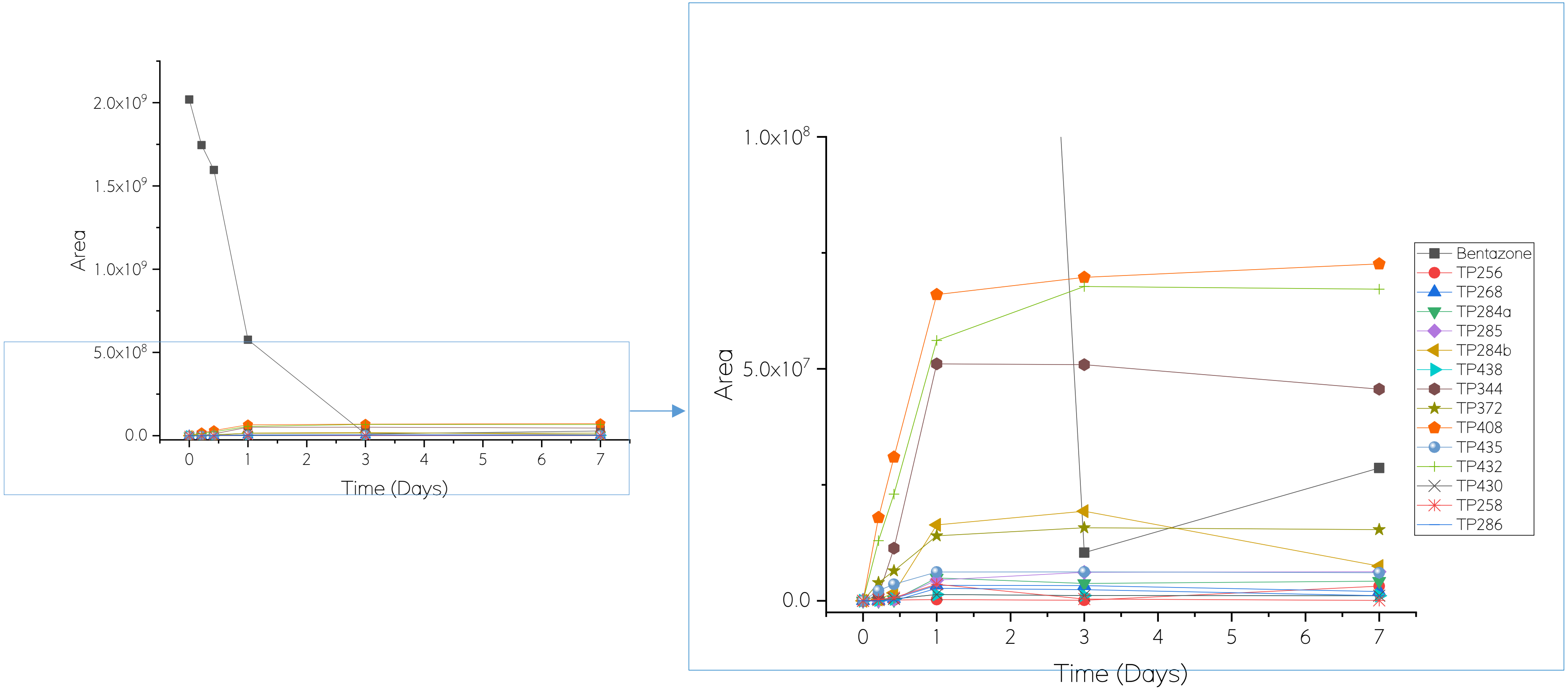

**Figure 6.** Evolution of bentazone and TPs during *T.versicolor* - mediated biodegradation experiments at lab-scale.

**Table 3.** Accurate mass measurements of the parent and fragment ions corresponding to bentazone TPs as determined by UPLC-(HESI-)-HRMS/MS, elemental composition with their theoretical *m/z*, relative mass error, and ring and double bond equivalents (RDB).

| Compound | $t_R$ (min) | *m/z* measured $[M-H]^-$ | Elemental composition | *m/z* theoretical $[M-H]^-$ | Relative mass error (ppm) | RDB |
|---|---|---|---|---|---|---|
| Bentazone | 5.10 | *239.0486** | $C_{10}H_{11}N_2O_3S$ | *239.0496* | *0.651* | *6.5* |
| | | 197.0014 | $C_7H_5N_2O_3S$ | 197.0021 | -0.707 | 6.5 |
| | | 175.0863 | $C_{10}H11N_2O$ | 175.0877 | -1.654 | 6.5 |
| | | 132.0313 | $C_7H_4N_2O$ | 132.0329 | -3.895 | 7.0 |
| TP256 (OH-BTZ) | 4.62 | *255.0439** | $C_{10}H_{11}N_2O_4S$ | *255.0445* | *-2.356* | *6.5* |
| | | 197.0015 | $C_7H_5N_2O_3S$ | 197.0021 | -0.707 | 6.5 |
| | | 191.0814 | $C_{10}H_{11}N_2O_2$ | 191.0826 | -5.709 | 6.5 |
| | | 132.0314 | $C_7H_4N_2O$ | 132.0329 | -3.895 | 7.0 |
| | | 79.9557 | $O_3S$ | 79.9574 | -21.173 | 1.0 |
| TP268 | 5.32 | *267.0440** | $C_{11}H_{11}N_2O_4S$ | *267.0445* | *2.382* | *7.5* |
| | | 224.9967 | $C_8H_5N_2O_4S$ | 224.9970 | 0.961 | 7.5 |
| | | 203.0815 | $C_{11}H_{11}N_2O_2$ | 203.0826 | 0.127 | 7.5 |
| | | 160.0265 | $C_8H_4N_2O_2$ | 160.0278 | -1.680 | 8.0 |
| TP285 | 5.84 | *284.0342** | $C_{10}H_{10}N_3O_5S$ | *284.0347* | *-1.494* | *7.5* |
| | | 241.9869 | $C_7H_4N_3O_5S$ | 241.9872 | -3.365 | 7.5 |
| | | 220.0719 | $C_{10}H_{10}N_3O_3$ | 220.0728 | -4.292 | 7.5 |
| | | 197.0014 | $C_7H_5N_2O_3S$ | 197.0021 | -0.707 | 6.5 |
| | | 177.0169 | $C_7H_3N_3O_3$ | 177.0180 | -6.210 | 8.0 |
| | | 132.0311 | $C_7H_4N_2O$ | 132.0329 | -3.895 | 7.0 |
| TP284a | 2.76 | *283.0389** | $C_{11}H_{11}N_2O_5S$ | *283.0394* | *-1.892* | *7.5* |
| | | 240.9915 | $C_8H_5N_2O_5S$ | 240.9919 | -3.549 | 7.5 |
| | | 239.0485 | $C_{10}H_{11}N_2O_3S$ | 239.0496 | -4.502 | 6.5 |
| | | 219.0765 | $C_{11}H_{11}N_2O_3$ | 219.0775 | -4.681 | 7.5 |
| | | 197.0014 | $C_7H_5N_2O_3S$ | 197.0021 | -0.707 | 6.5 |
| | | 177.0286 | $C_8H_5N_2O_3$ | 177.0300 | -11.328 | 7.5 |
| | | 176.0210 | $C_8H_4N_2O_3$ | 176.0227 | -10.001 | 8 |
| TP284b | 5.67 | *283.0388** | $C_{11}H_{11}N_2O_5S$ | *283.0394* | *-2.104* | *7.5* |
| | | 239.0487 | $C_{10}H_{11}N_2O_3S$ | 239.0496 | -3.875 | 6.5 |
| | | 197.0014 | $C_7H_5N_2O_3S$ | 197.0021 | -0.707 | 6.5 |
| | | 175.0863 | $C_{10}H_{11}N_2O$ | 175.0877 | -1.654 | 6.5 |
| | | 133.0393 | $C_7H_5N_2O$ | 133.0402 | -10.720 | 6.5 |
| | | 132.0313 | $C_7H_4N_2O$ | 132.0329 | -12.581 | 7.0 |
| | | 59.0115 | $C_2H_3O_2$ | 59.0133 | -39.190 | 1.5 |
| TP438 | 5.55 | *437.0781** | | | | |
| | | 239.0486 | $C_{10}H_{11}N_2O_3S$ | 239.0496 | 0.651 | 6.5 |
| | | 197.0012 | $C_7H_5N_2O_3S$ | 197.0021 | -0.707 | 6.5 |
| | | 175.0862 | $C_{10}H11N_2O$ | 175.0877 | -1.654 | 6.5 |
| TP344 | 5.96 | *343.0748** | | | | |
| TP372 | 5.84 | *371.0913** | | | | |
| | | 239.0487 | $C_{10}H_{11}N_2O_3S$ | 239.0496 | 0.651 | 6.5 |
| | | 197.0015 | $C_7H_5N_2O_3S$ | 197.0021 | -0.707 | 6.5 |
| | | 175.0863 | $C_{10}H11N_2O$ | 175.0877 | -1.654 | 6.5 |
| | | 160.0389 | | | | |
| | | 116.0489 | | | | |
| TP408 | 5.83 | *407.0676** | | | | |
| | | 239.0488 | $C_{10}H_{11}N_2O_3S$ | 239.0496 | 0.651 | 6.5 |
| | | 175.0864 | $C_{10}H11N_2O$ | 175.0877 | -1.654 | 6.5 |
| | | 132.0312 | $C_7H_4N_2O$ | 132.0329 | -3.895 | 7.0 |
| TP435 | 5.85 | *434.0869** | | | | |
| | | 241.0442 | | | | |
| | | 239.0487 | $C_{10}H_{11}N_2O_3S$ | 239.0496 | 0.651 | 6.5 |
| | | 198.9963 | | | | |
| | | 175.0866 | $C_{10}$H11$N_2O$ | 175.0877 | -1.654 | 6.5 |
| | | 116.0490 | | | | |
| | | 61.9867 | | | | |
| TP432 | 5.85 | *431.1922** | | | | |
| | | 239.0488 | $C_{10}H_{11}N_2O_3S$ | 239.0496 | 0.651 | 6.5 |
| | | 175.0863 | $C_{10}$H11$N_2O$ | 175.0877 | -1.654 | 6.5 |
| | | 132.0312 | $C_7H_4N_2O$ | 132.0329 | -3.895 | 7.0 |
| | | 59.0122 | $C_2H_3O_2$ | 59.0133 | | |
| TP430 | 5.92 | *429.2486** | | | | |
| | | 239.0490 | $C_{10}H_{11}N_2O_3S$ | 239.0496 | 0.651 | 6.5 |
| | | 167.0308 | | | | |
| | | 59.0116 | $C_2H_3O_2$ | 59.0133 | -20.435 | 1.5 |
| TP258 | 5.39 | *257.0230** | $C_9H_9N_2O_5S$ | 257.0238 | 0.745 | 6.5 |
| | | 214.9756 | $C_6H_3N_2O_5S$ | 214.9763 | -0.551 | 6.5 |
| | | 193.0604 | $C_9H_9N_2O_3$ | 193.0619 | -0.511 | 6.5 |
| | | 151.0134 | $C_6H_3N_2O_2$ | 151.0144 | -2.771 | 6.5 |
| | | 134.0105 | $C_6H_2N_2O_2$ | 134.0116 | -3.797 | 7.0 |
| TP286 | 4.64 | *285.0182** | $C_{10}H_9N_2O_6S$ | 285.0187 | 2.234 | 7.5 |
| | | 257.0232 | $C_9H_9N_2O_5S$ | 257.0238 | 1.912 | 6.5 |
| | | 214.9759 | $C_6H_3N_2O_5S$ | 214.9763 | 1.016 | 6.5 |
| | | 193.0610 | $C_9H_9N_2O_3$ | 193.0619 | 1.301 | 6.5 |
| | | 151.0136 | $C_6H_3N_2O_2$ | 151.0144 | -1.072 | 6.5 |
| | | 150.0055 | $C_6H_2N_2O_2$ | 150.0071 | -3.132 | 7.0 |
| | | 135.0182 | $C_6H_2N_2O_2$ | 135.0195 | -5.213 | 6.5 |
| | | 77.9638 | $NO_2S$ | 77.9650 | -7.7483 | 1.5 |
| TP220 | 3.15 | *219.0070** | | | | |
| | | 191.0119 | | | | |
| | | 163.0168 | | | | |
| | | 121.9537 | | | | |
| | | 97.9866 | | | | |
| | | 77.9637 | $NO_2S$ | 77.9650 | -8.729 | 1.5 |
| TP388 | 6.25 | *387.1225** | | | | |
| | | 239.0489 | $C_{10}H_{11}N_2O_3S$ | 239.0496 | 0.651 | 6.5 |
| | | 141.0154 | | | | |
| | | 116.9710 | | | | |
| | | 59.0116 | $C_2H_3O_2$ | 59.0133 | -20.604 | 1.5 |
| TP518 | 5.85 | *517.1489** | | | | |
| TP192 | 4.54 | *191.0119** | | | | |
| | | 163.0168 | | | | |
| | | 121.9537 | | | | |
| | | 59.0121 | $C_2H_3O_2$ | 59.0133 | -10.606 | 1.5 |
| TP494 | 5.85 | *493.0856** | $C_{20}H_{21}N_4O_7S_2$ | 493.0857 | -0.165 | 12.5 |

*corresponds to the *precursor ion*

### 3.4. Removal of bentazone in TBR bioreactor

Once proved that *T. versicolor* was able to degrade bentazone, its degrading ability in time under non-sterile conditions was evaluated. For this, a TBR containing wood chips immobilized with the fungus was set up, through which fortified agricultural wastewater was treated, and no additional carbon sources were added. The TBR was operated in the SBR mode because information on its performance can be obtained in a short period, instead of waiting for the stationary-state as in continuous flow treatments. The operation of the TBR in SBR mode presents additional advantages over the purely continuous mode, such as higher removal efficiencies, and less energy consumption. However, the SBR system also presents some limitations such as the lack of automatization and consequently, the requirement of high maintenance during the operation. The bioreactor stability was assessed along with the operation. A 3-day cycle was arbitrarily established, ensuring that bentazone was not completely removed at the end of each batch so that the degradation performance of the TBR could be evaluated.

In previous experiences with TBR [35], high adsorption of the investigated pollutants onto the wood chips was observed. The adsorption rate is highly dependent on the physical-chemical properties of the pollutant. Thus, to evaluate the adsorption rate of bentazone on TBR, two reactors were set up: one containing pine wood chips colonized with the fungus (experimental TBR) and another one filled only with lignocellulosic material (control TBR). The results obtained using a recirculating flow of 70 mL $min^{-1}$ are shown in *Figure 7*. The laccase activity, measured throughout the treatment, reached a maximum value of 46.68 AU $L^{-1}$ during the first batch (*Figure 7a*). This could be explained by its accumulation within the static incubation. Then, the enzymatic activity kept at a low constant level and showed a decreasing trend, probably ascribed to the fungus aging behavior and biomass washing out.

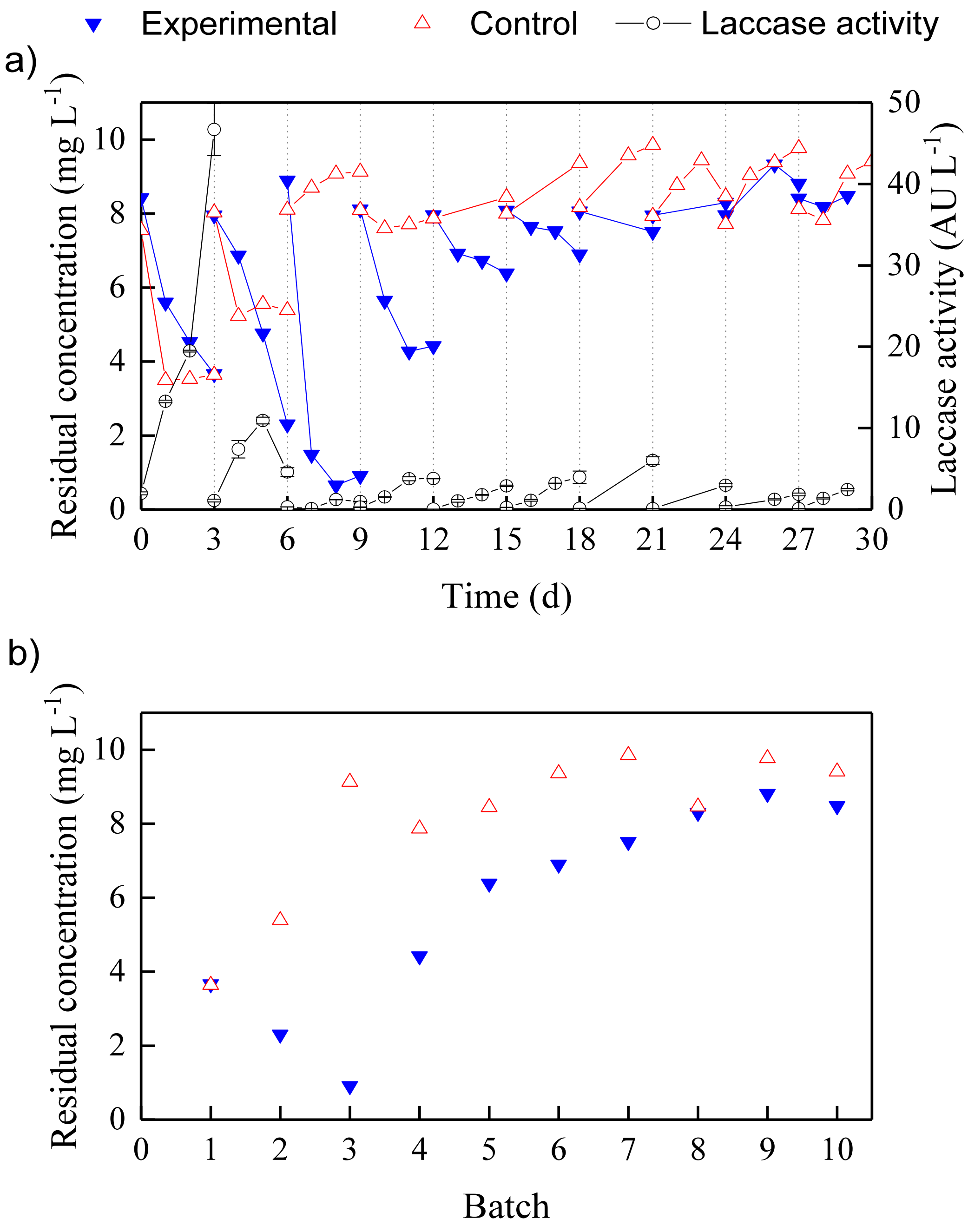

**Figure 7.** Removal of bentazone by *T. versicolor* in sequencing batch TBR under non-sterile conditions. a. Time-course of bentazone variation during the treatment; b. The final concentration of bentazone in each batch.

In comparison with the introduced amount, bentazone removal in the TBR was moderate (48% on average), and lower than that observed in lab-scale batch reactors. This could be explained by the less amount of biomass in the TBR (approximately 4.2 x $10^{-3}$ g d.w. biomass/g dry wood, equivalent to 0.84 g d.w. $L^{-1}$) than in the Erlenmeyer reactors (2.5 g d.w. $L^{-1}$), the short operation time fixed for the TBR and the less available nutrients in the reactor (pine wood chips in the TBR vs the rich defined medium in the lab-scale batch reactors). Adsorption contributed to more than half (58%) of the removal, which is in line with the fact that the employed lignocellulosic materials serve as an effective sorbent for pesticide removal [48]. The final concentration of bentazone after each batch is shown in Figure 7b. As can be seen, adsorption decreased throughout the sequential batches, while the removal yield was always higher in the experimental TBR than in the control TBR. This suggests that the immobilized fungus maintained bioactive for 30 days. Thus, this result proves that the TBR tested, besides being very simple in terms of configuration, is highly cost-efficient, since there was neither addition of C nor N sources during the treatment to maintain the fungus activity, a reusable waste is used as support, and requires low operation energy. The highest expenses would result from the long retention times (3 days) and pH adjustment of the water.

The experimental reactor was stopped after one month of operation because its removal efficiency dropped to 29%, whereas the value in the control reactor was 11%. Although decreasing bioactivity was observed, the obtained results are still promising, considering that limited biomass was introduced without any other complementary nutrients. Hence, some improvements such as the replacement of the lignocellulosic support materials or scale-up of the reactor could be taken into consideration to increase the amount of immobilized biomass. Furthermore, the contact time between the immobilized fungus and the pesticide was quite low based on empirical calculations and therefore, the contribution from the cytochrome P450 system to bentazone degradation could be restricted.

HPC was monitored during the TBR treatment (*Table 4*). CFU count did not significantly increase during the treatment. It was indeed reduced within the first 4 batches compared to the original value (4.68 log CFU $mL^{-1}$, *Table 1*). This could be attributed to nutrient limitation. Thus this approach offers strong potential for dealing with bacterial contamination that represents one of the main barriers in the implementation of fungal reactors to treat wastewater [35]. The adjustment of the wastewater pH to 4.5 is also useful to favor the fungus bioactivity and limit bacteria growth [49]. Furthermore, fewer microbial counts were found in the experimental set than in the control. A reasonable hypothesis is that the more active the fungus is, the stronger the antagonistic interactions between the fungus and the bacteria may be [50]. COD was also analyzed during TBR operation, and it dropped in both sets during operation. Although the control reactor showed overall a lower COD than the experimental one, the COD content was still much higher than the original value (31.85 mg $L^{-1}$, *Table 1*). This could be explained by the addition of bentazone to the water, the elution of wood particles from the packed bed, and/or the wood rotting by *T. versicolor*, and needs to be addressed in future research.

**Table 4**. HPC of the agricultural wastewater throughout the sequencing batch TBR treatment

| Batch | Microbial counts [Log (CFU) $mL^{-1}$] | |
|---|---|---|
| | Experimental | Control |
| 2 | 2.90 | 6.28 |
| 4 | 3.90 | 5.99 |
| 6 | 6.10 | 5.80 |
| 8 | 5.99 | 6.59 |
| 10 | 5.87 | 5.71 |

## 4. Conclusions

*T. versicolor* could effectively degrade bentazone, during which both laccase and cytochrome P450 were involved. Up to 19 TPs were captured and identified, indicating that hydroxylations, oxidations, methylations, N-nitrosation, and dimerization played important roles during the detoxification process. The TBR system operated in SBR mode was effective to remove bentazone throughout 30 days of operation, and thus, represents a promising strategy to deal with bentazone contamination at a real scale. However, some improvements should be considered in future research to address the high COD levels resulting in water and the low biomass present in the reactor. The operation of the TBR without supplementing nutrients and at acidic pH values aids in its good performance under non-sterile conditions. While the latter would result in additional expenses during the implementation of the process at real-scale, they could be offset by the no-nutrients requirement. Overall, this work points out *T. versicolor* as a suitable candidate towards bentazone degradation and the adopted reactor system shows promise for bentazone bioremediation at a real scale.

## Acknowledgments

This work has been supported by the Spanish Ministry of Economy and Competitiveness State Research Agency (CTM2016-75587-C2-1-R and CTM2016-75587-C2-2-R) and co-financed by the European Union through the European Regional Development Fund (ERDF) and the Horizon 2020 research and innovation WATERPROTECT project (727450). This work was partly supported by the Generalitat de Catalunya (Consolidate Research Group 2017-SGR-01404) and the Ministry of Science and Innovation (Project CEX2018-000794-S). The Department of Chemical, Biological and Environmental Engineering of the Universitat Autònoma de Barcelona is a member of the Xarxa de Referència en Biotecnologia de la Generalitat de Catalunya. K. Hu acknowledges the financial support from the Chinese Scholarship Council.

**Conflicts of interest**

The authors declare that no conflict of interest exists in the submission of this manuscript.

## Supporting Information

# Remediation of bentazone contaminated water by *Trametes versicolor*: characterization, identification of transformation products, and implementation in a trickle-bed reactor under non-sterile conditions

*Manuel García-Vara [1], Kaidi Hu [2], Cristina Postigo [1*], Lluc Olmo [2], Gloria Caminal [3], Montserrat Sarrà [2], Miren López de Alda [1*]*

[1]*Water, Environmental and Food Chemistry Unit (ENFOCHEM), Department of Environmental Chemistry, Institute of Environmental Assessment and Water Research (IDAEA), Spanish Council for Scientific Research (CSIC), Jordi Girona 18-26, 08034 Barcelona, Spain*

[2]*Departament d'Enginyeria Química, Biològica i Ambiental, Escola d'Enginyeria, Universitat Autònoma de Barcelona, 08193 Bellaterra, Barcelona, Spain*

[3]*Institut de Química Avançada de Catalunya (IQAC), CSIC, Jordi Girona 18-26, 08034 Barcelona, Spain*

[*]***Corresponding authors:***

***Cristina Postigo, Miren López de Alda***

E-mail: cprqam@cid.csic.es, mlaqam@cid.csic.es

Institute of Environmental Assessment and Water Research (IDAEA-CSIC)

Department of Environmental Chemistry

C/ Jordi Girona 18-26, 08034 Barcelona, Spain.

Tel: +34-934-006-100, Fax: +34-932-045-904,

**Number of pages:** 22

**Number of figures:** 19

## List of figures

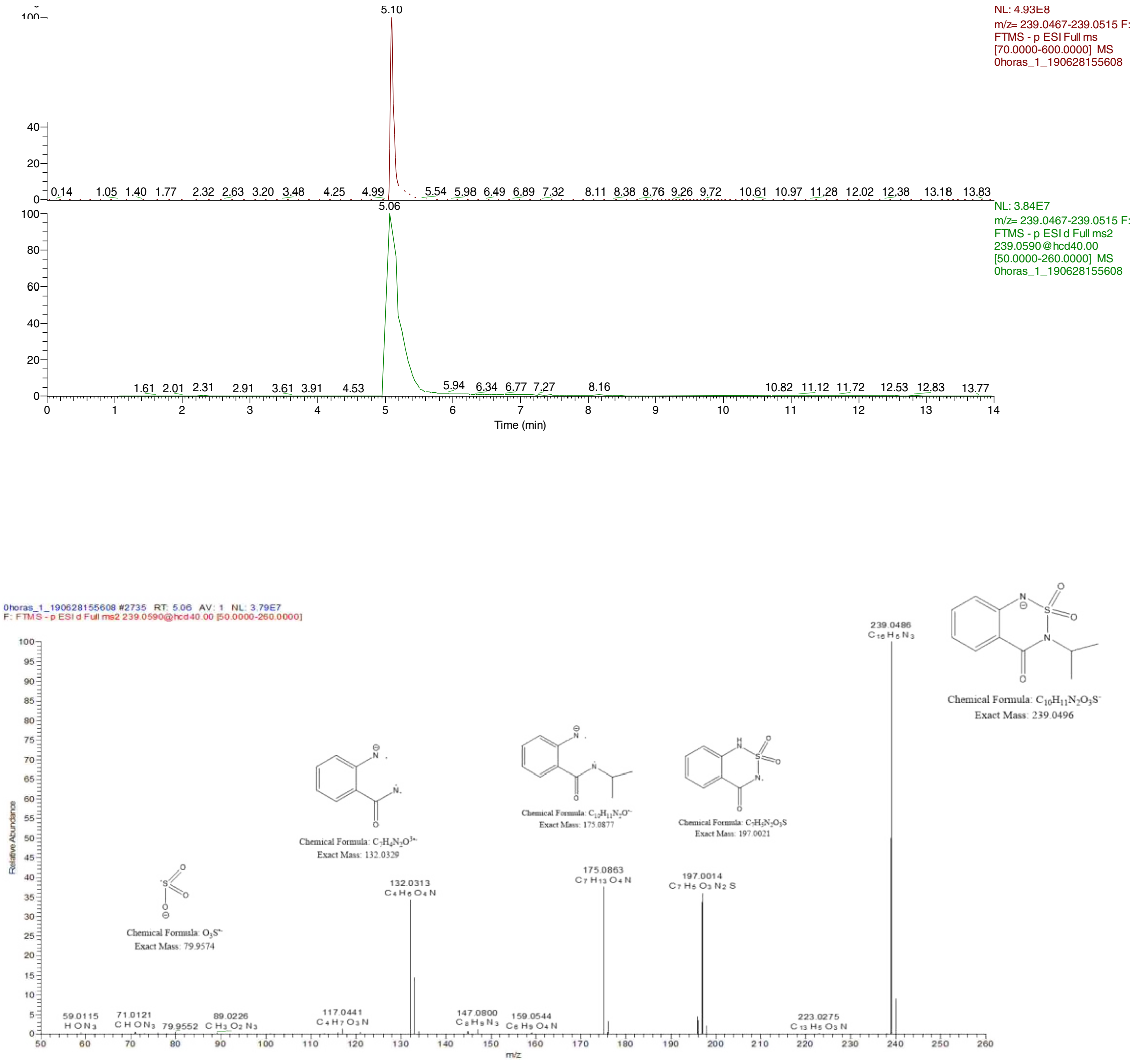

**Figure S1.** Extracted ion chromatograms (top: full MS, down: full MS$^2$) and MS$^2$ spectrum of *m/z* 239.0496 (bentazone).

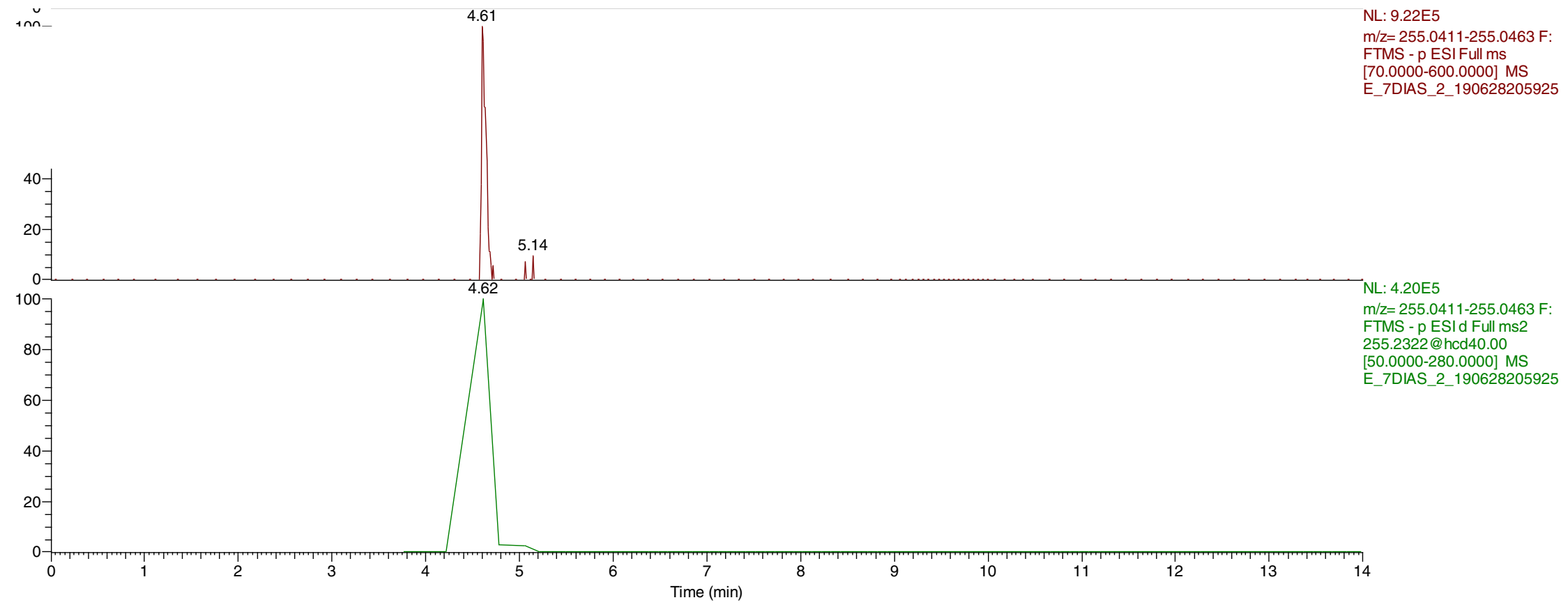

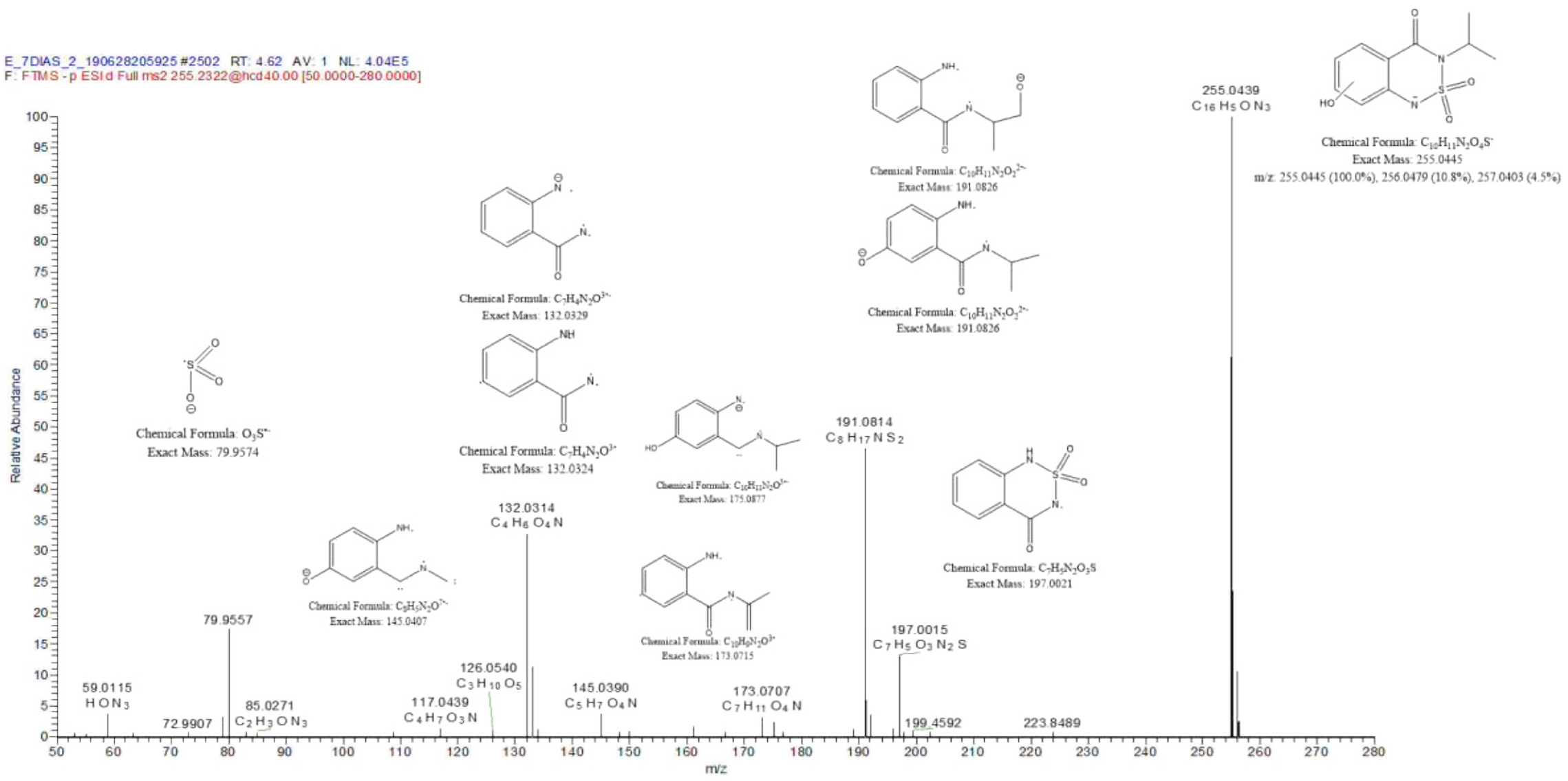

**Figure S2.** Extracted ion chromatograms (top: full MS, down: full MS$^2$) and MS$^2$ spectrum of *m/z* 255.0445 (TP256, OH-BTZ).

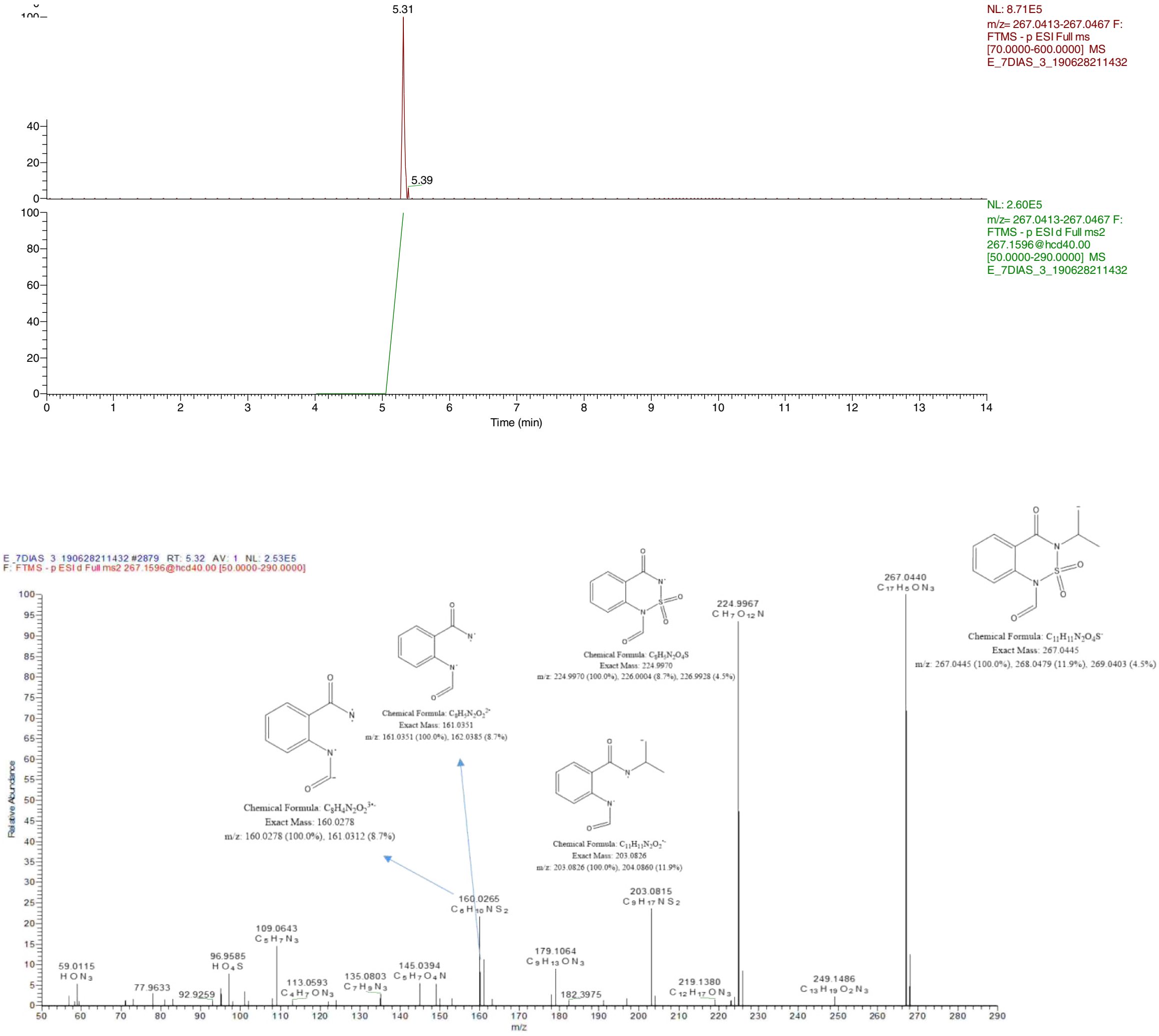

**Figure S3.** Extracted ion chromatograms (top: full MS, down: full MS$^2$) and MS$^2$ spectrum of *m/z* 267.0445 (TP268).

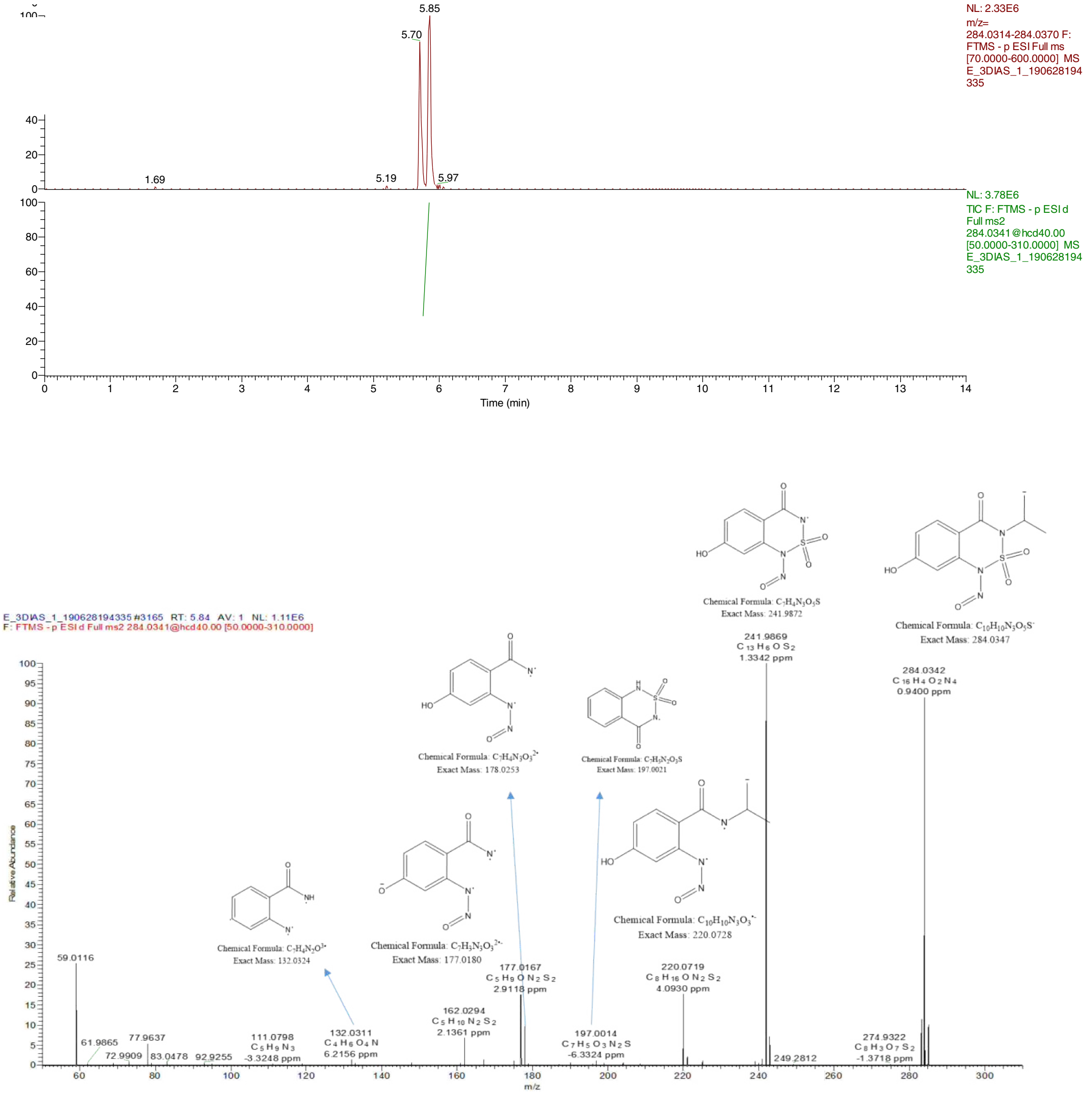

**Figure S4.** Extracted ion chromatograms (top: full MS, down: full MS$^2$) and MS$^2$ spectrum of *m/z* 284.0347 (TP285).

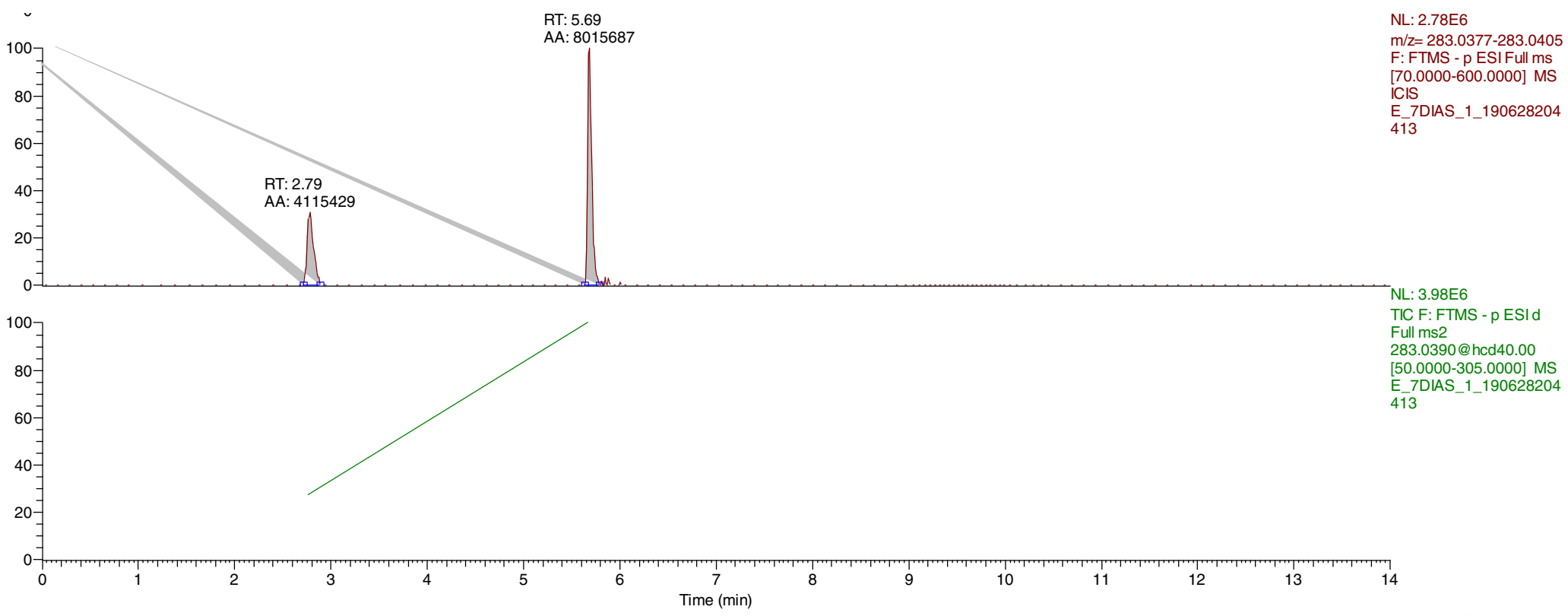

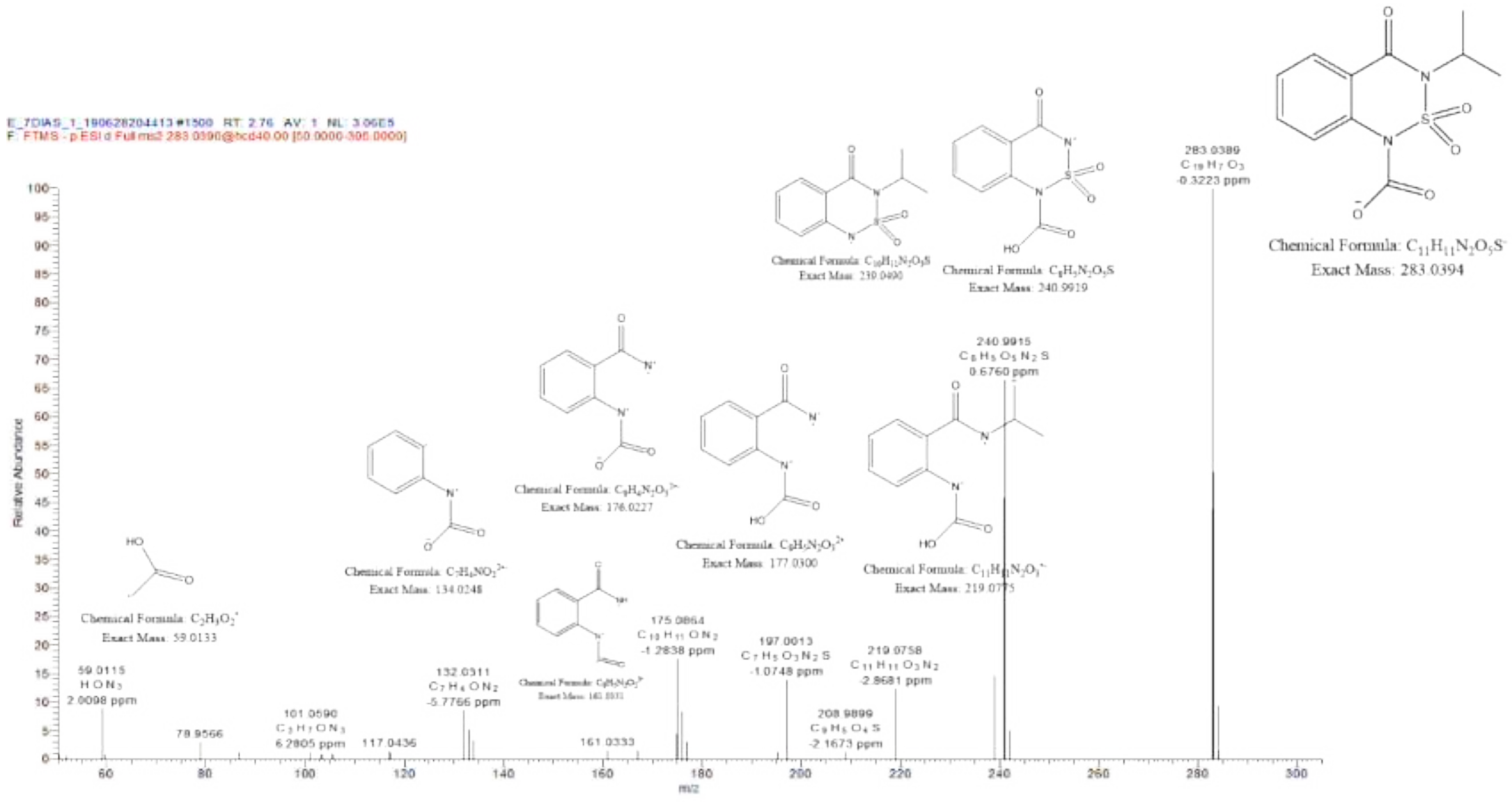

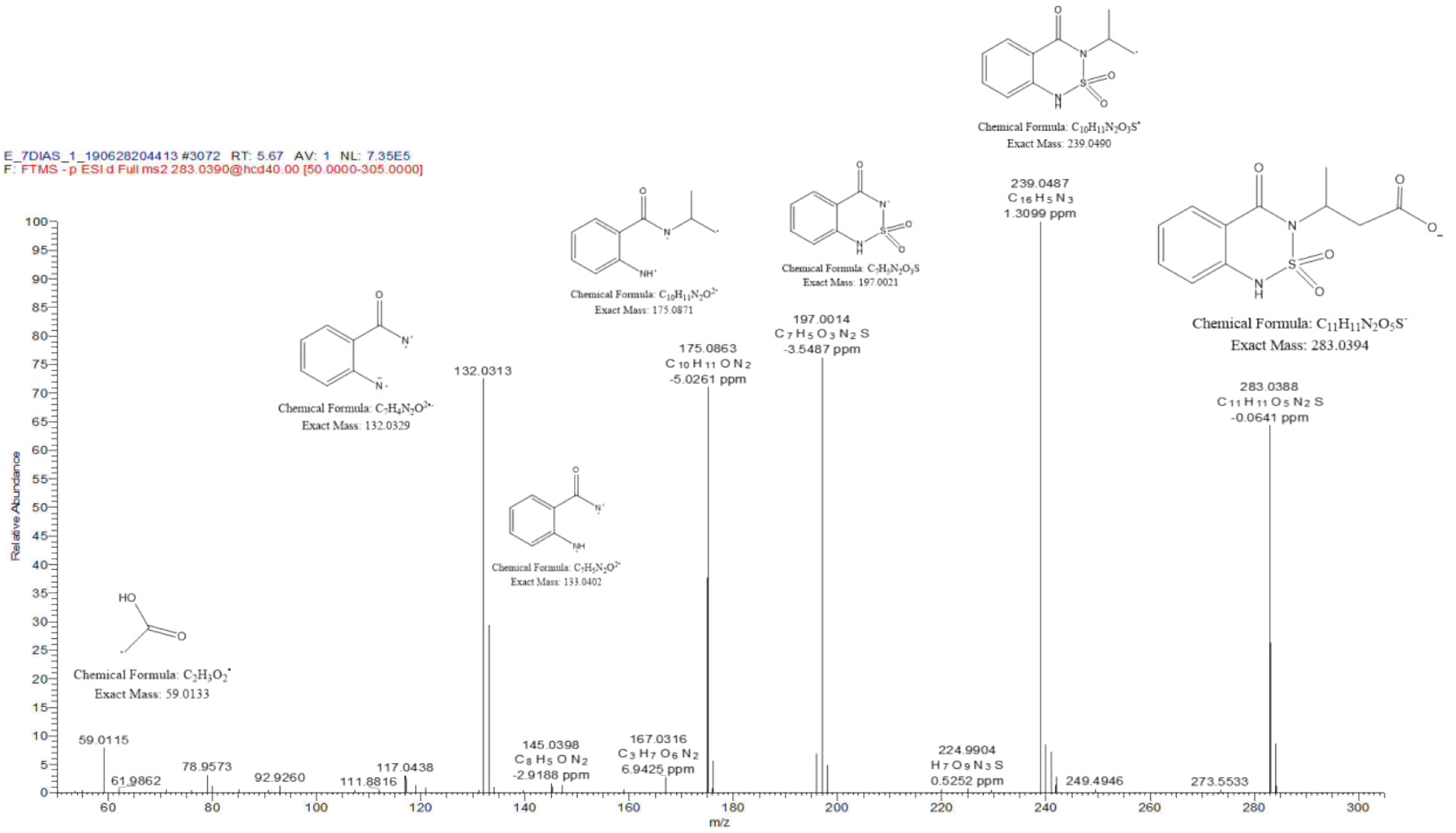

**Figure S5.** Extracted ion chromatograms (top: full MS, down: full MS$^2$) and MS$^2$ spectra of *m/z* 283.0394 at $t_R$ 2.76min and 5.67 min (TP284a and TP284b, respectively).

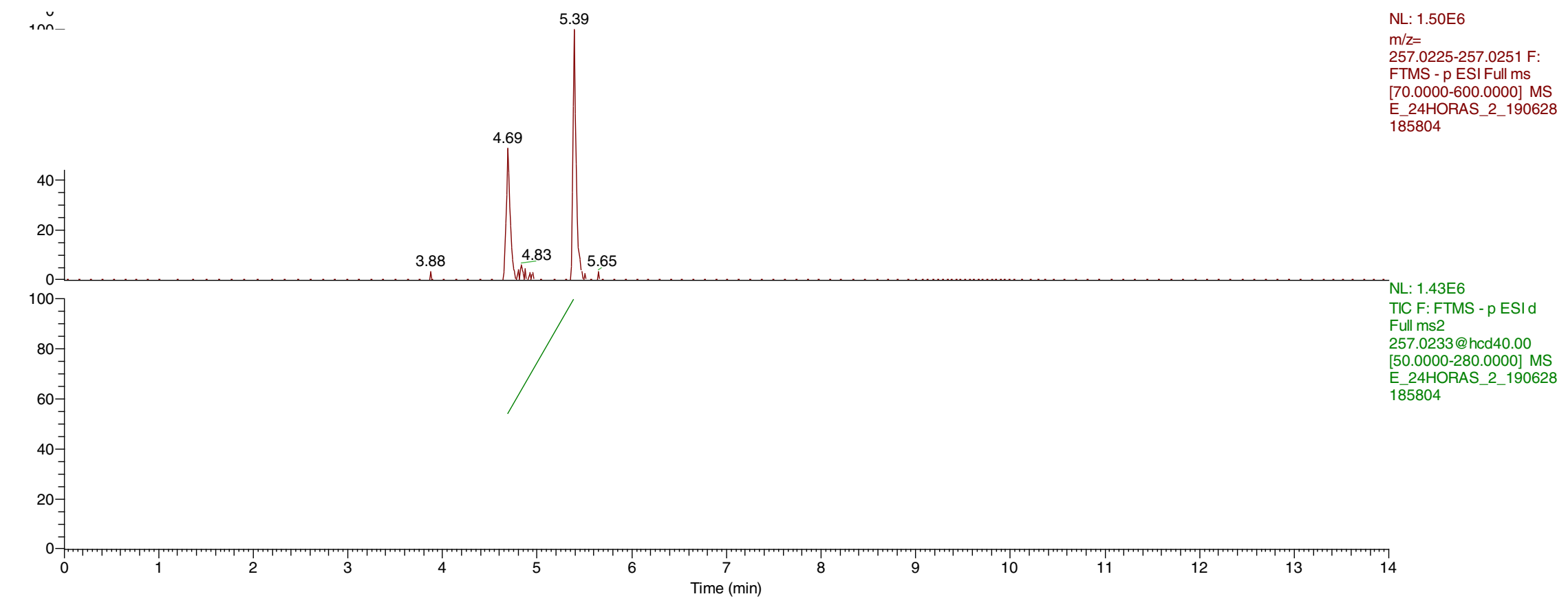

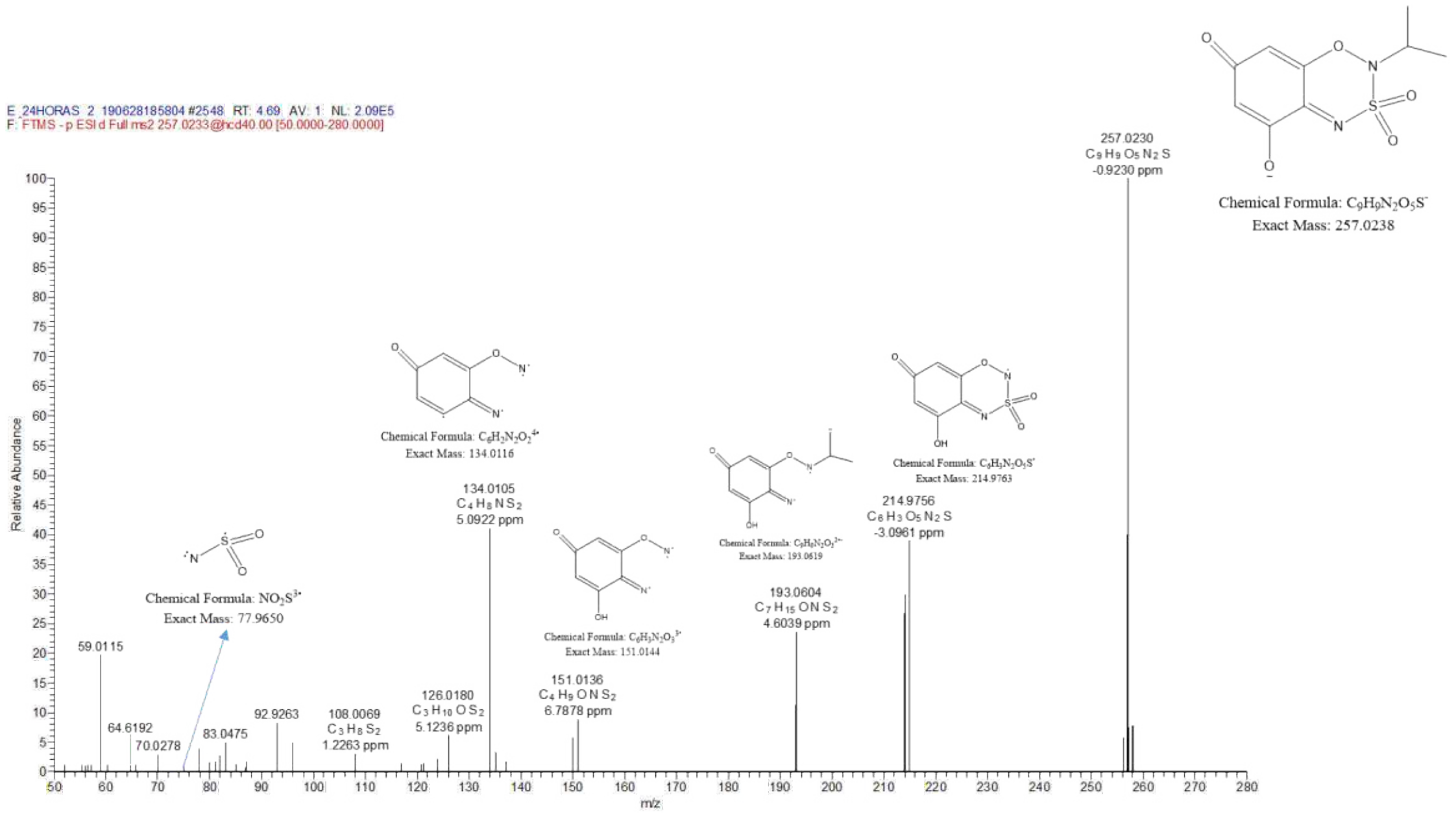

**Figure S6.** Extracted ion chromatograms (top: full MS, down: full MS$^2$) and MS$^2$ spectrum of *m/z* 257.0238 (TP258).

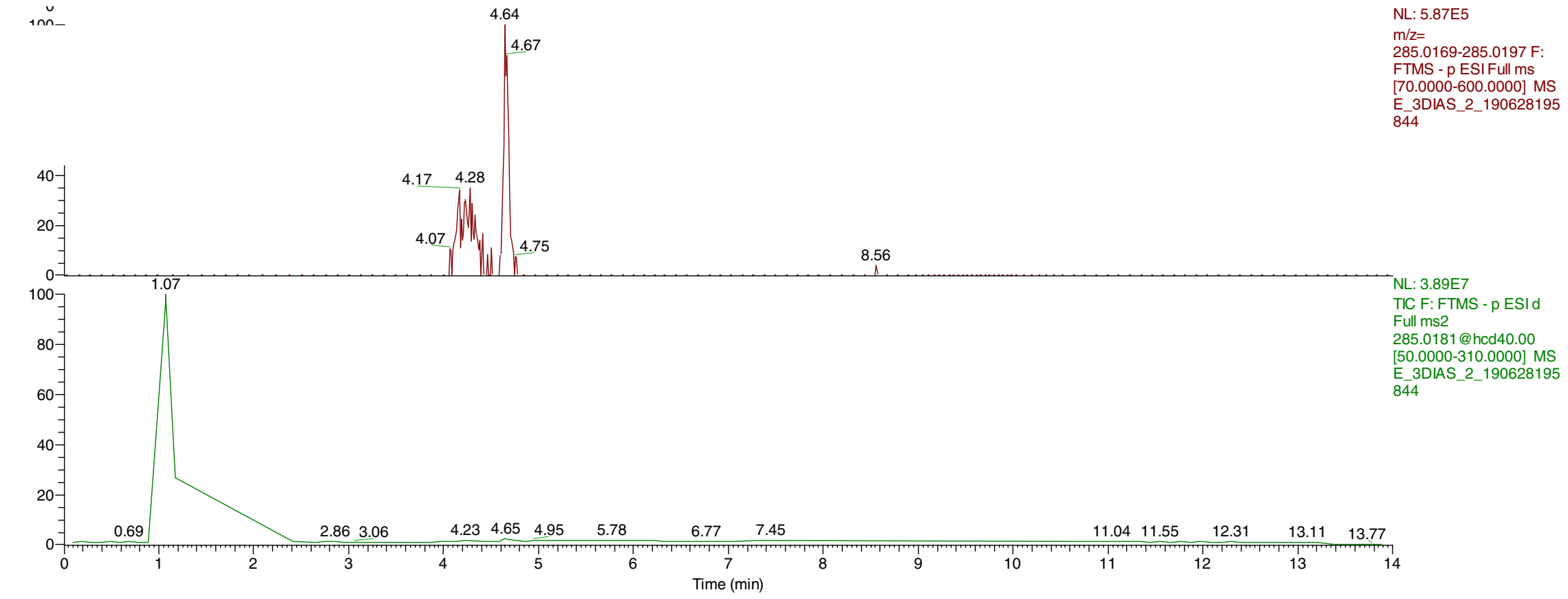

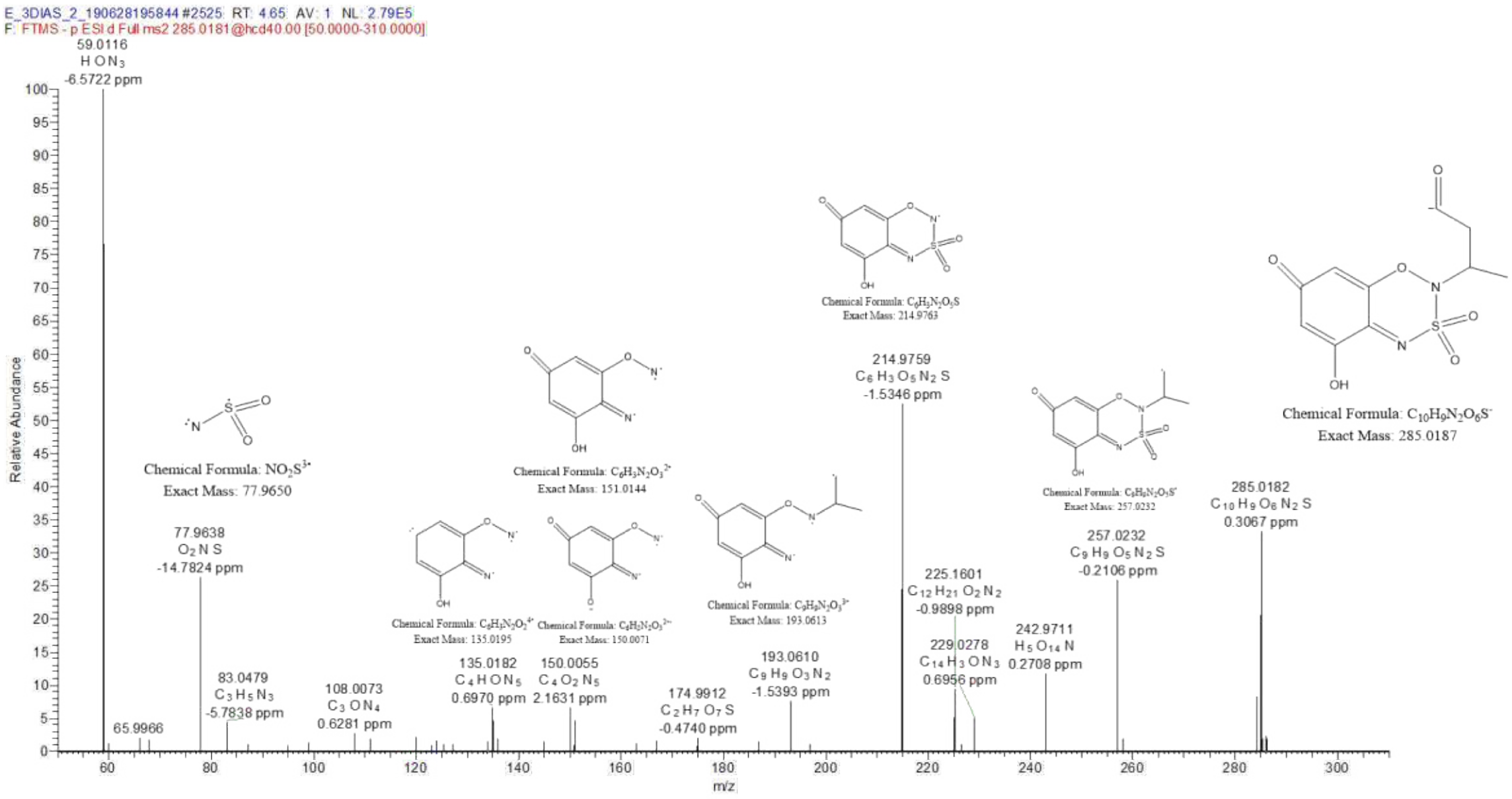

**Figure S7.** Extracted ion chromatograms (top: full MS, down: full $MS^2$) and $MS^2$ spectrum of *m/z* 285.0187 (TP286).

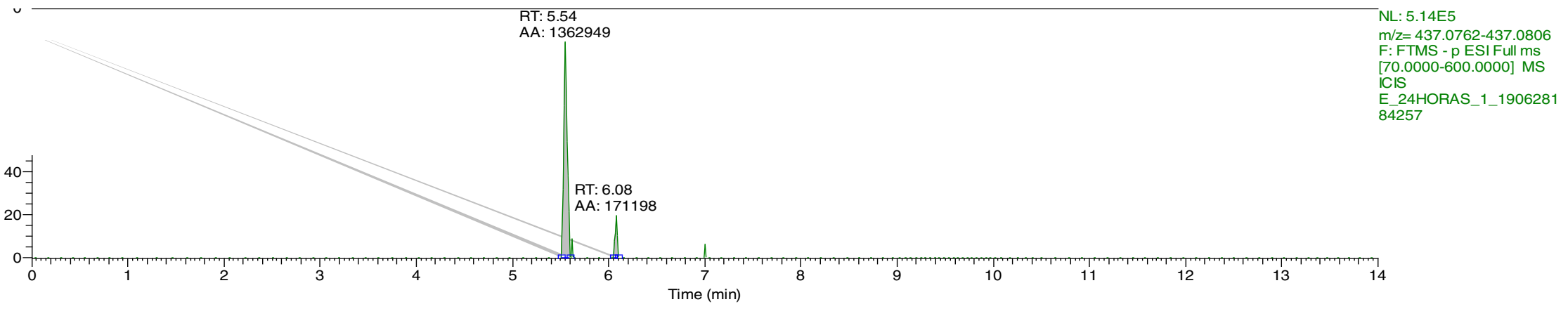

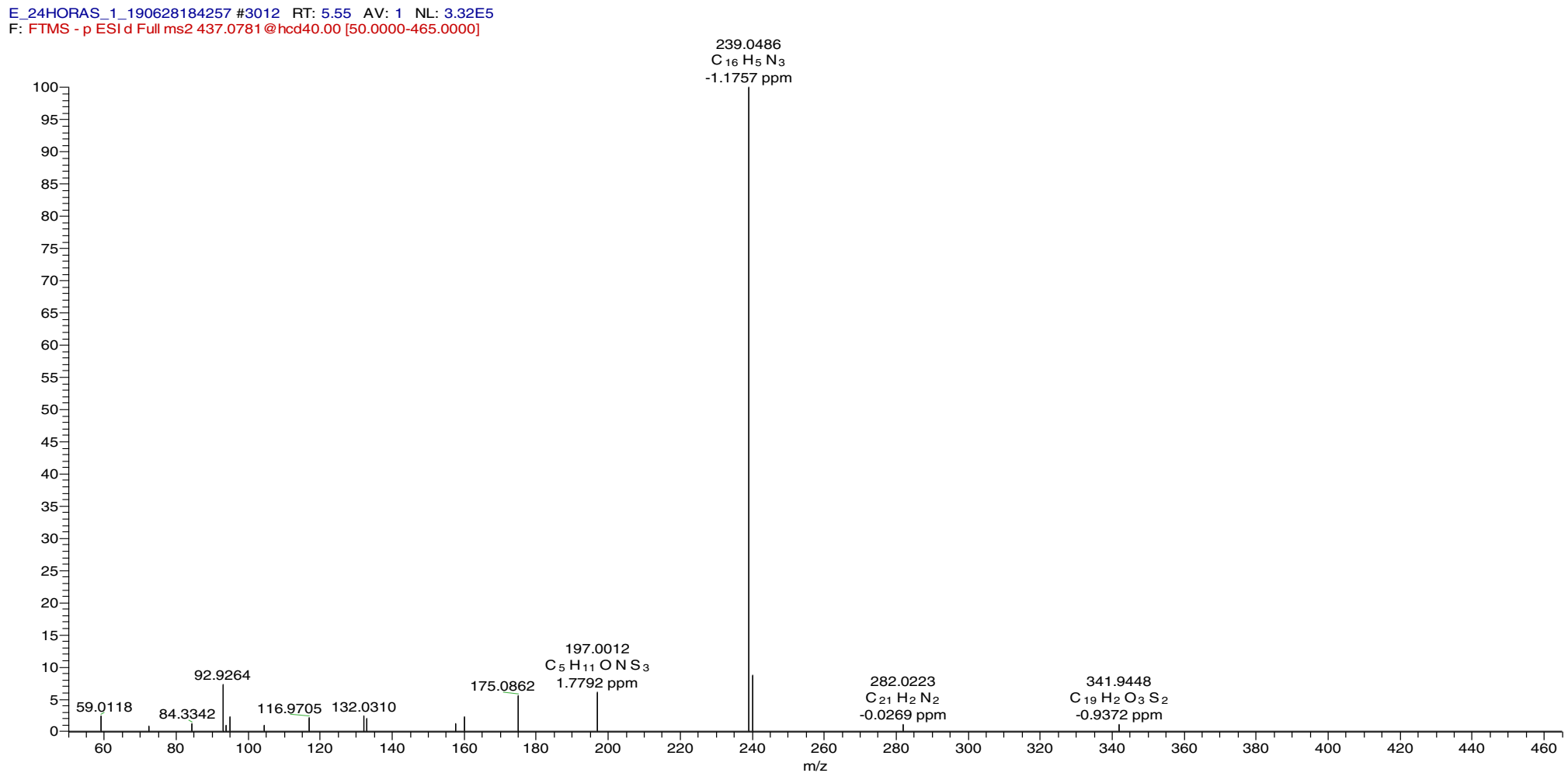

**Figure S8.** Extracted ion chromatogram and MS$^2$ spectrum of *m/z* 437.0781 (TP438).

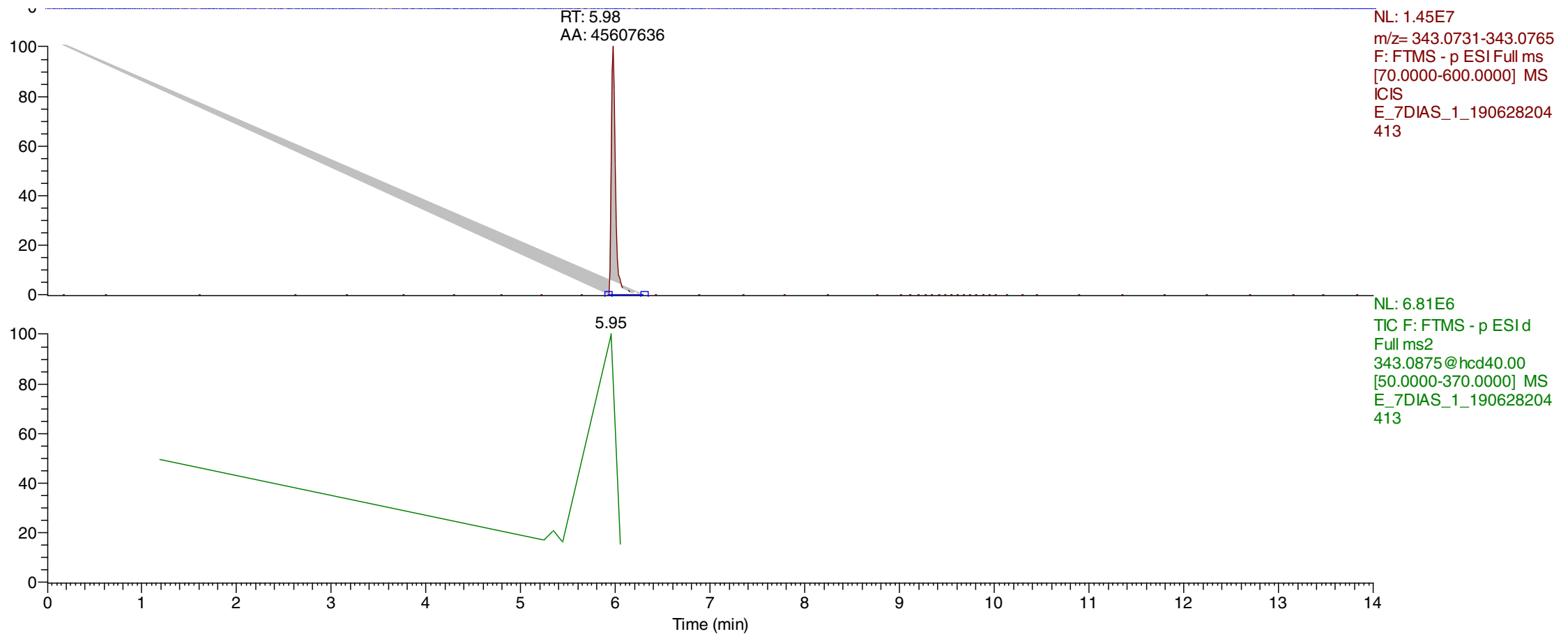

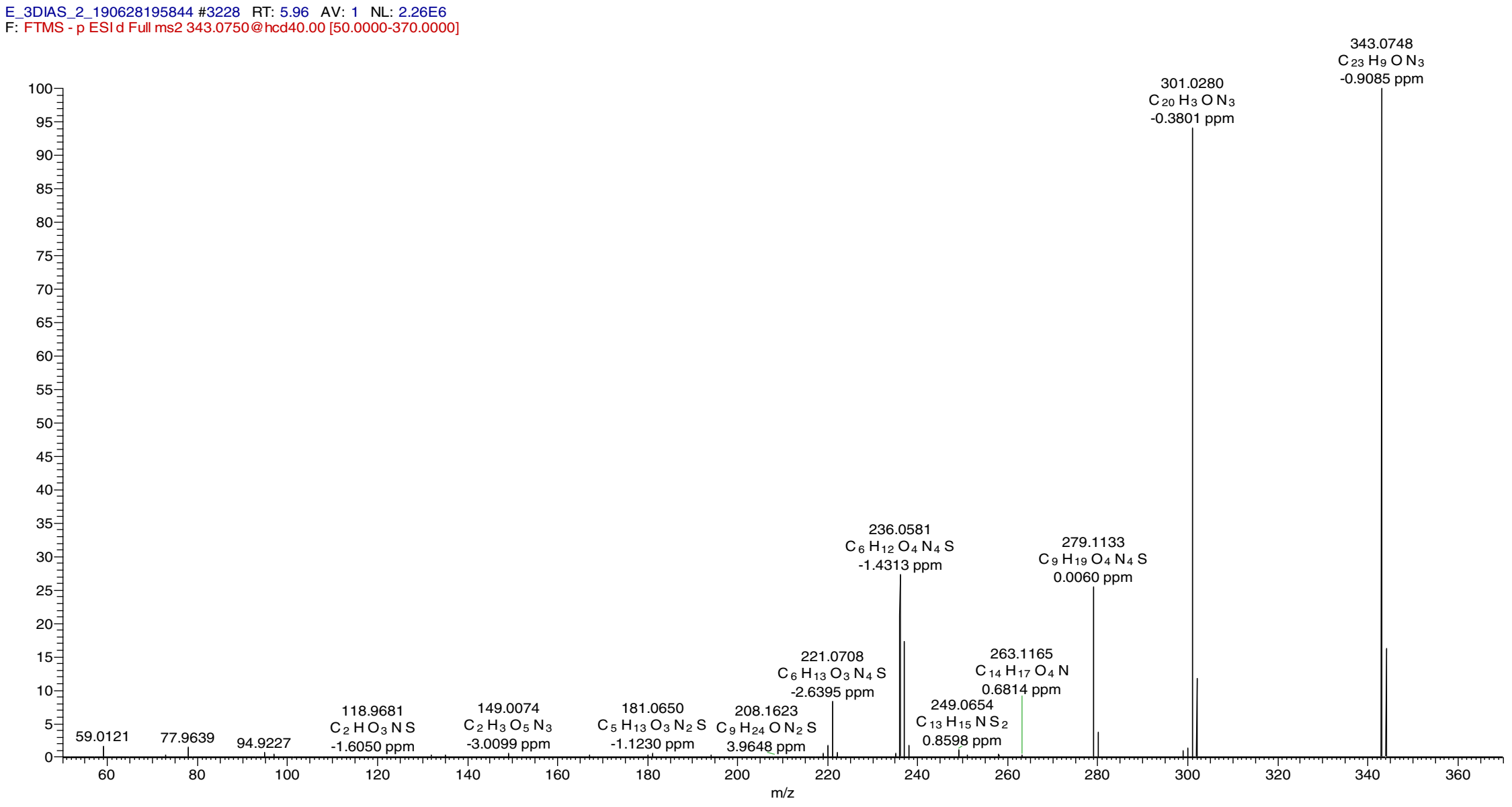

**Figure S9.** Extracted ion chromatograms (top: full MS, down: full MS$^2$) and MS$^2$ spectrum of *m/z* 343.0748 (TP344).

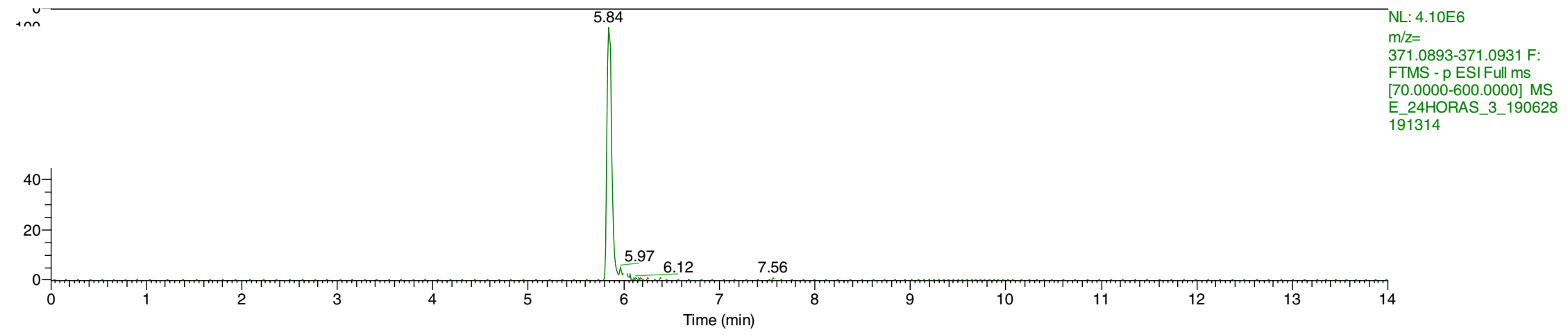

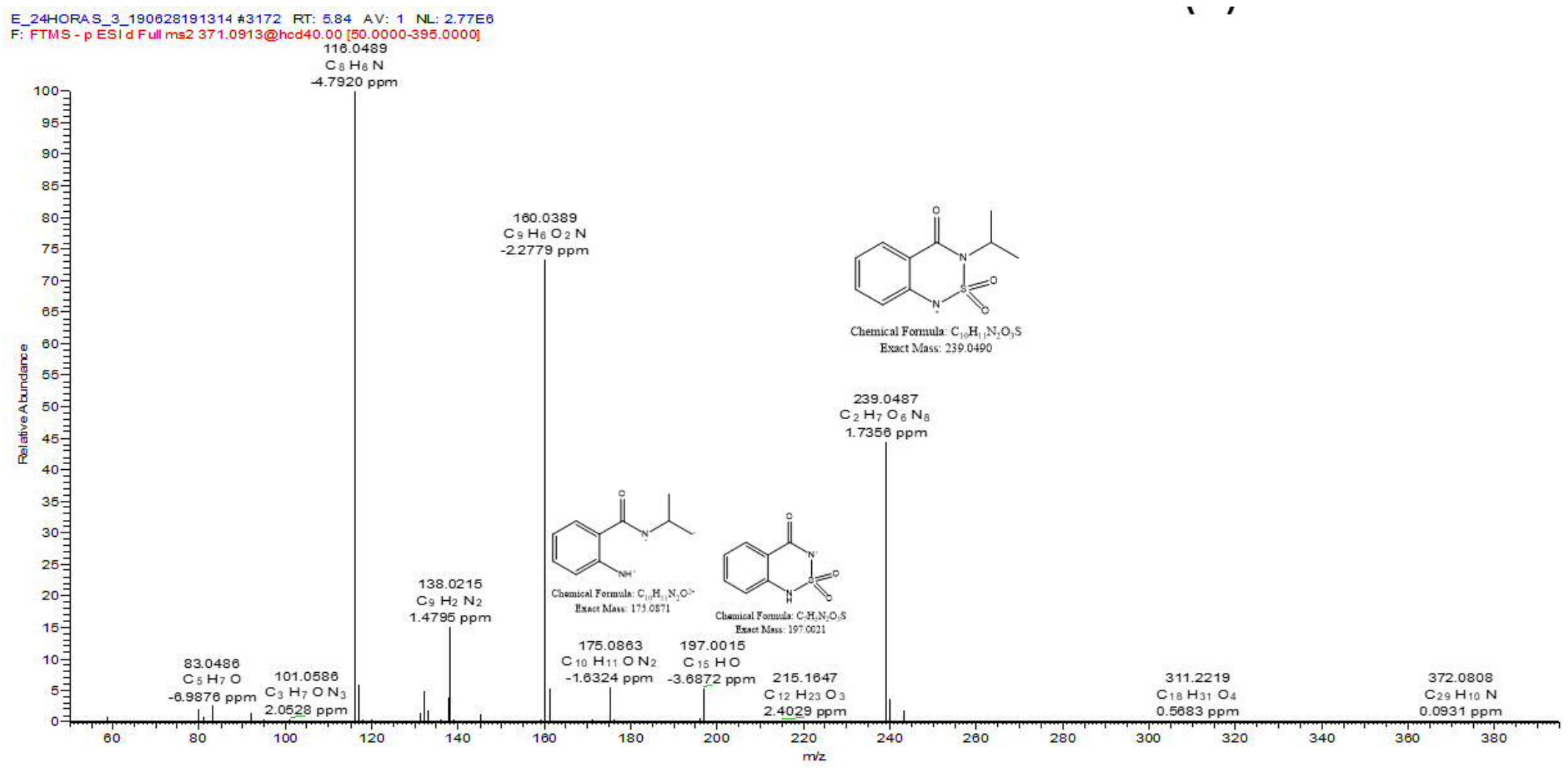

**Figure S10.** Extracted ion chromatogram and $MS^2$ spectrum of *m/z* 371.0913 (TP372).

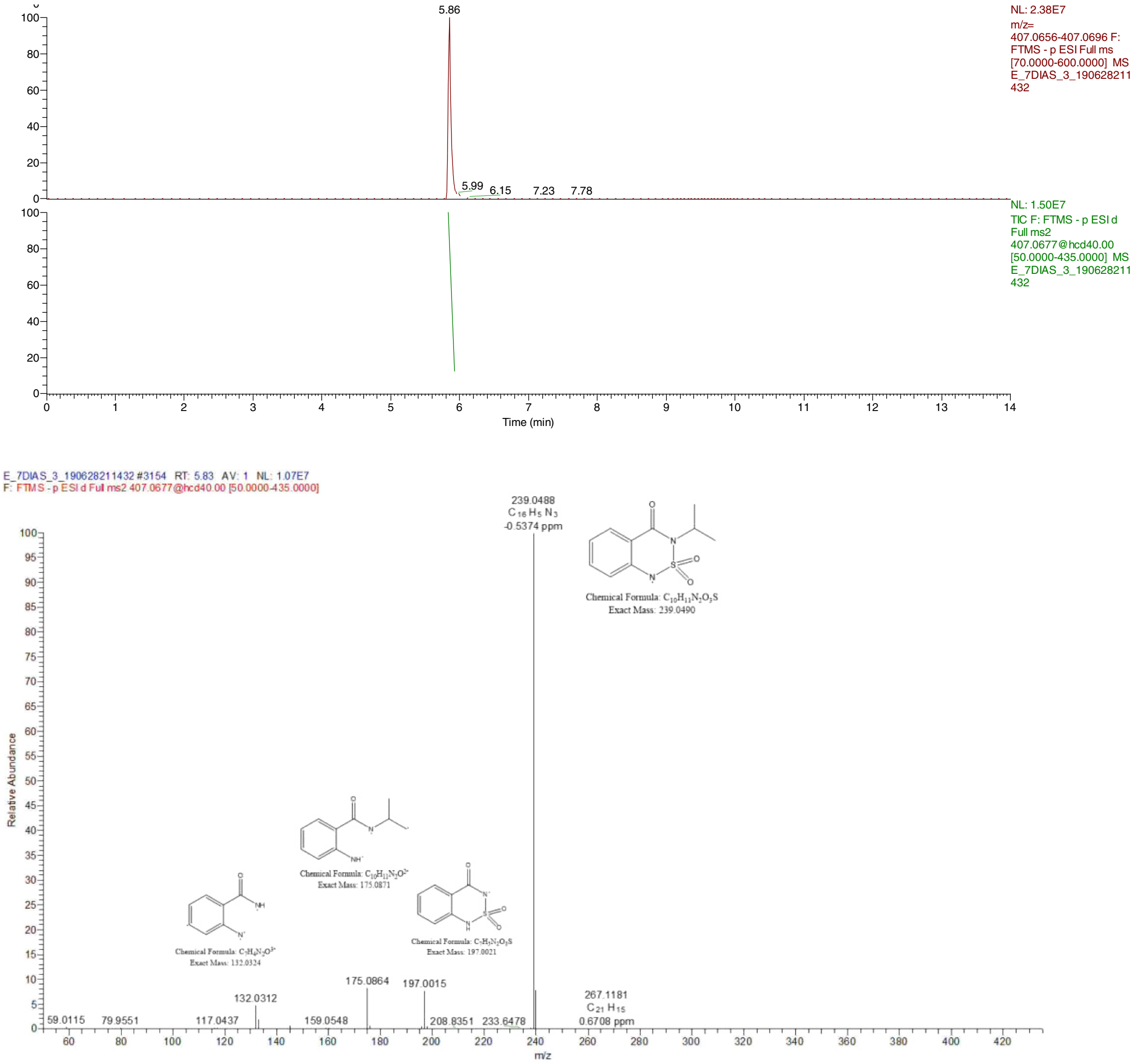

**Figure S11.** Extracted ion chromatograms (top: full MS, down: full MS$^2$) and MS$^2$ spectrum of *m/z* 407.0676 (TP408).

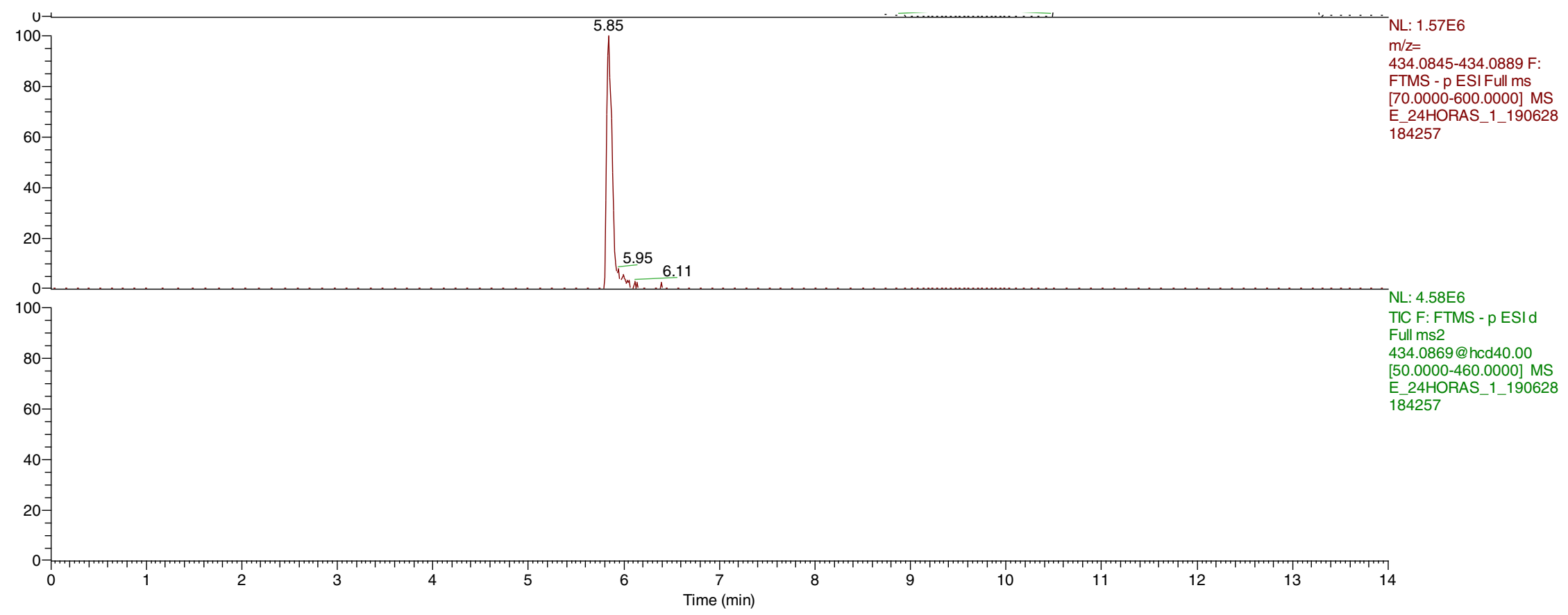

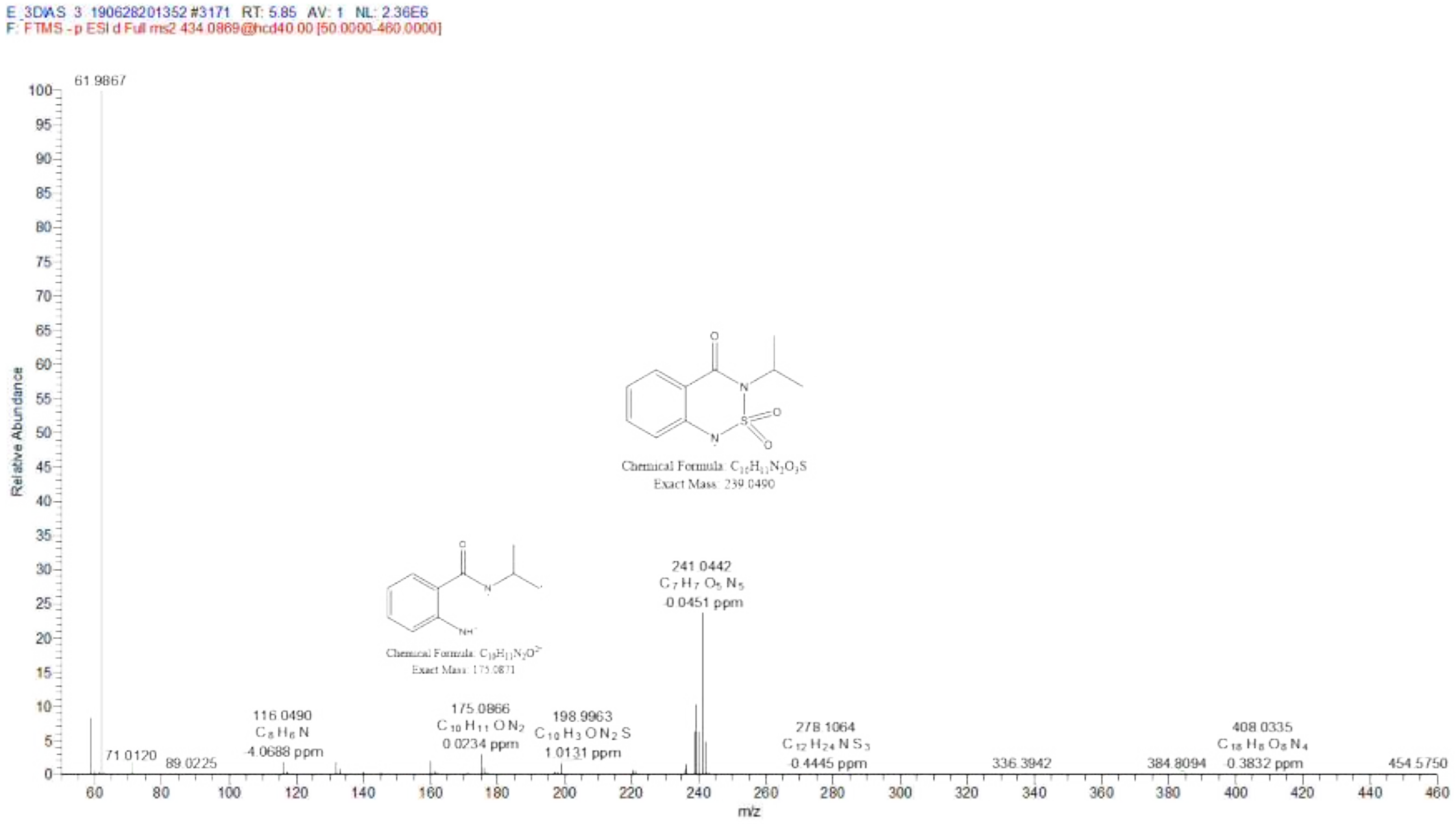

**Figure S12.** Extracted ion chromatograms (top: full MS, down: full MS$^2$) and MS$^2$ spectrum of *m/z* 434.0869 (TP435).

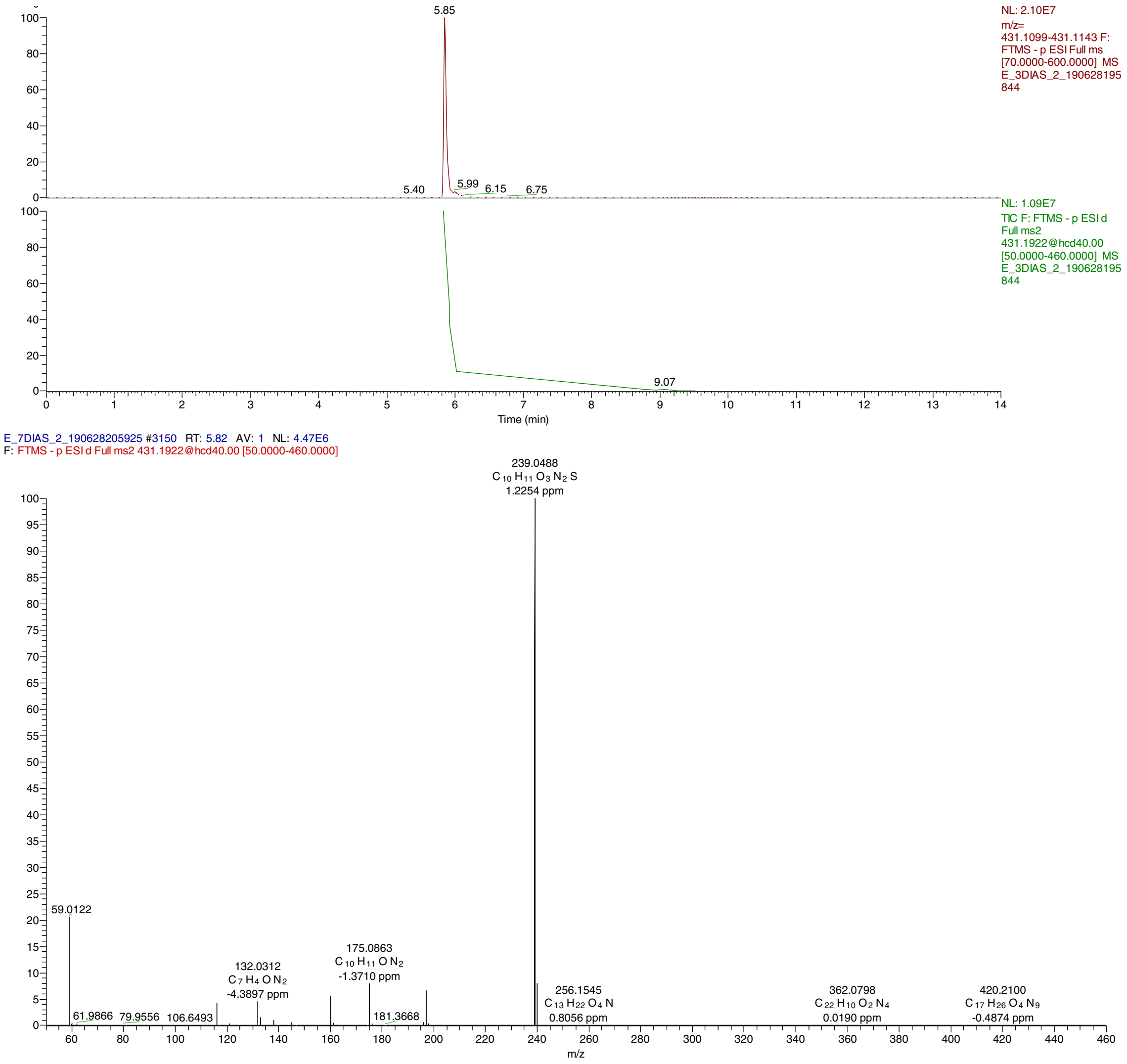

**Figure S13**. Extracted ion chromatograms (top: full MS, down: full MS$^2$) and MS$^2$ spectrum of *m/z* 431.1922 (TP432).

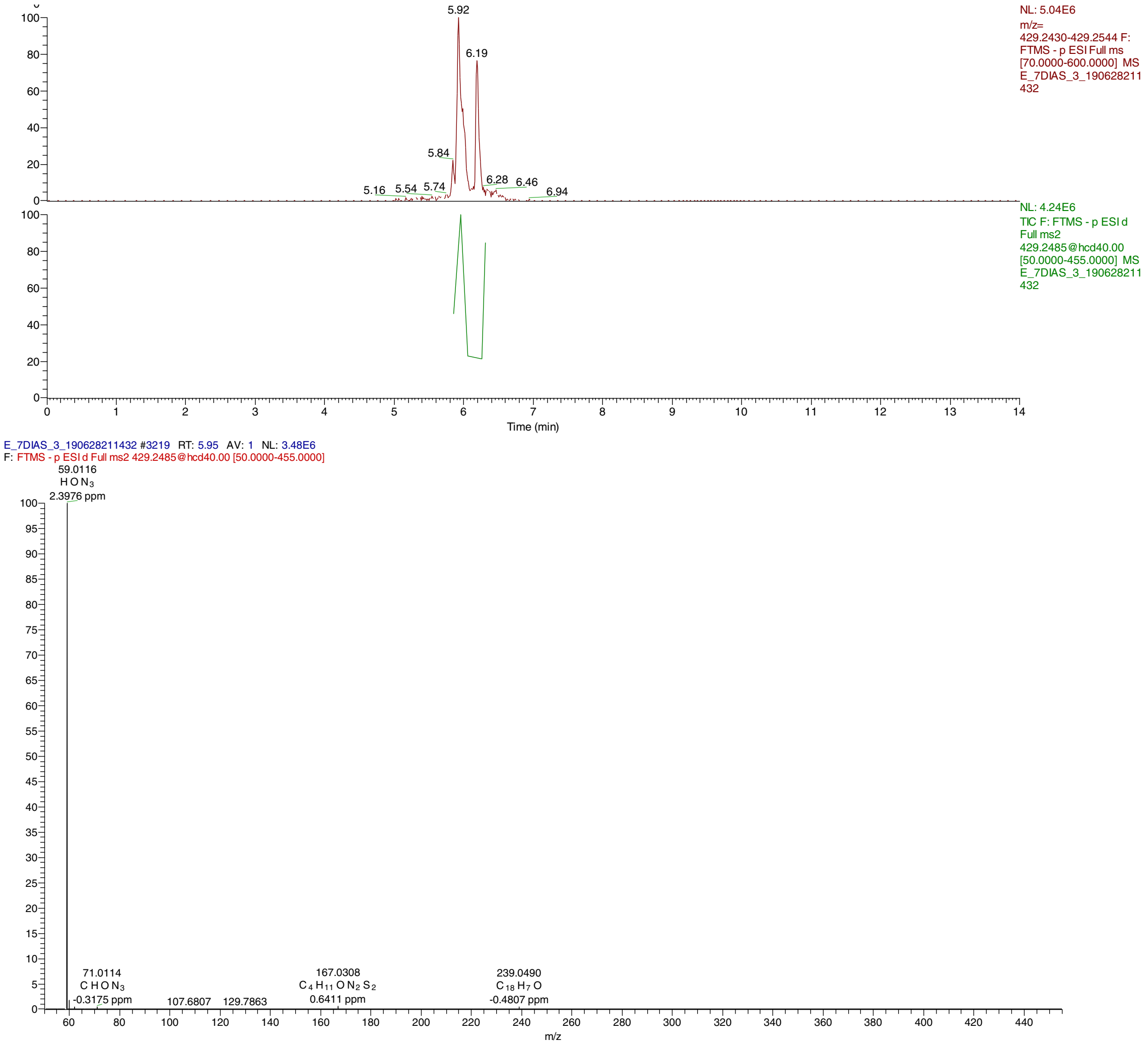

**Figure S14.** Extracted ion chromatograms (top: full MS, down: full MS$^2$) and MS$^2$ spectrum of *m/z* 429.2486 (TP430).

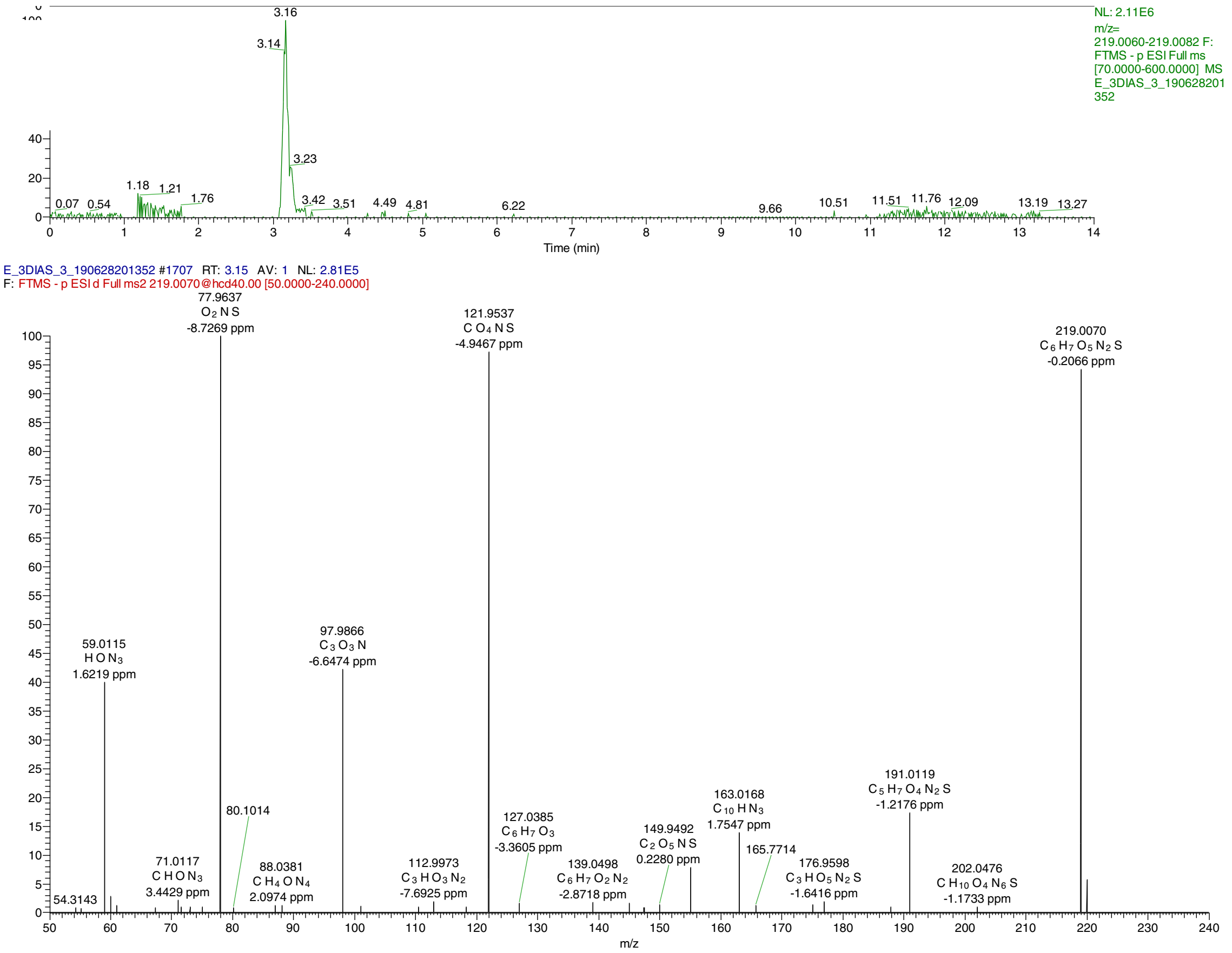

**Figure S15.** Extracted ion chromatogram and MS$^2$ spectrum of *m/z* 219.0070 (TP220).

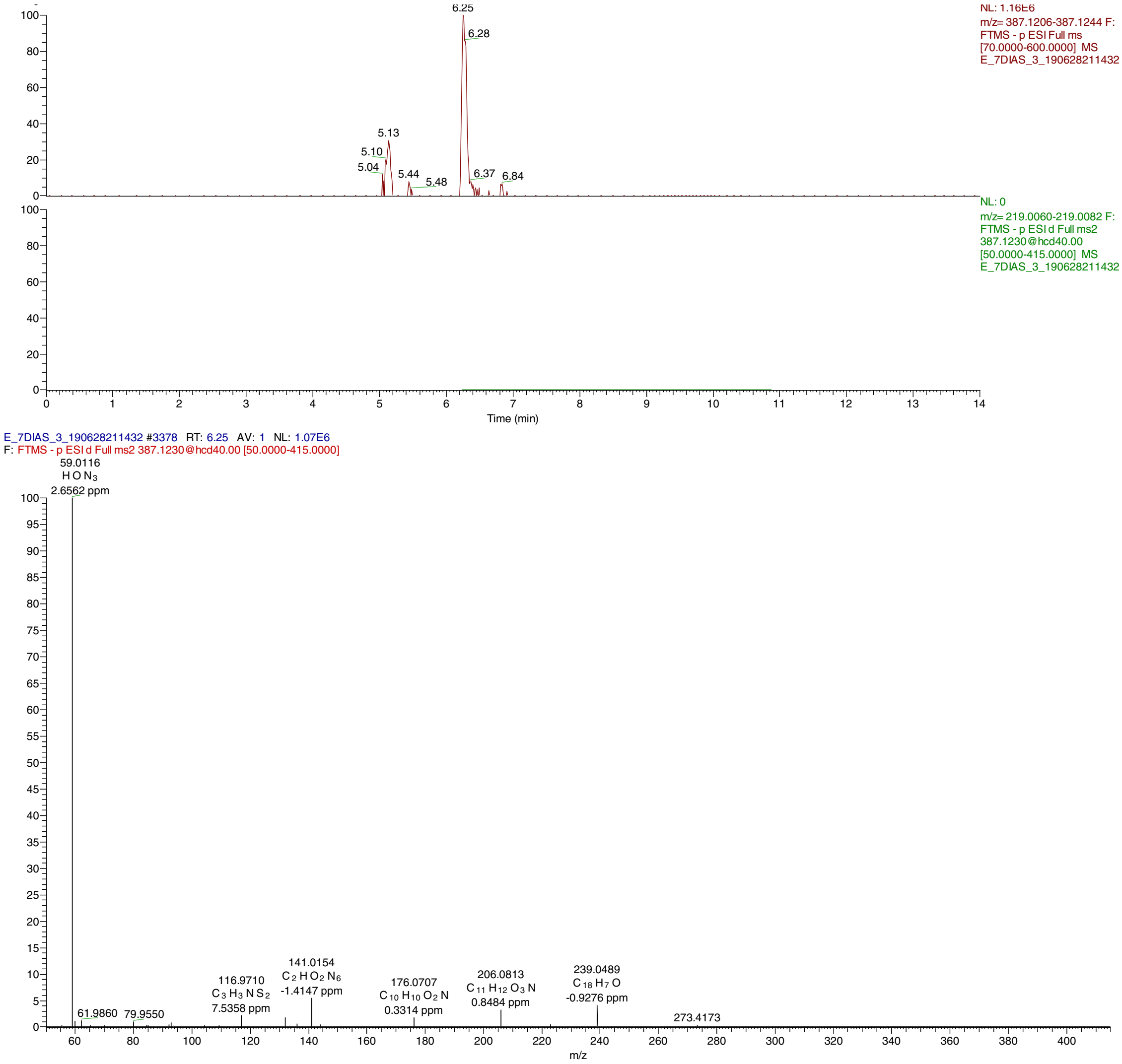

**Figure S16.** Extracted ion chromatograms (top: full MS, down: full MS$^2$) and MS$^2$ spectrum of *m/z* 387.1225 (TP388).

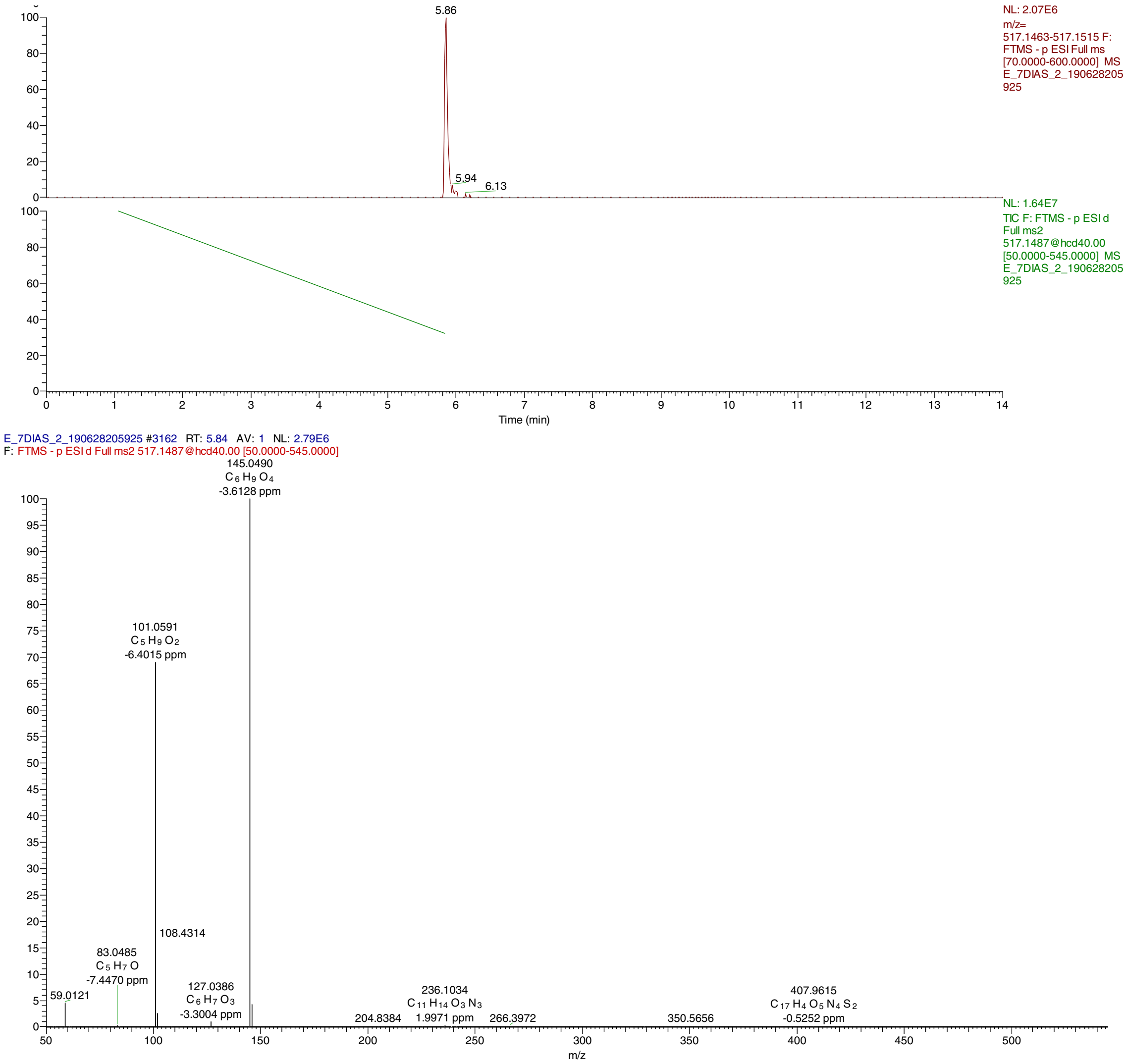

**Figure S17.** Extracted ion chromatograms (top: full MS, down: full MS$^2$) and MS$^2$ spectrum of *m/z* 517.1489 (TP518).

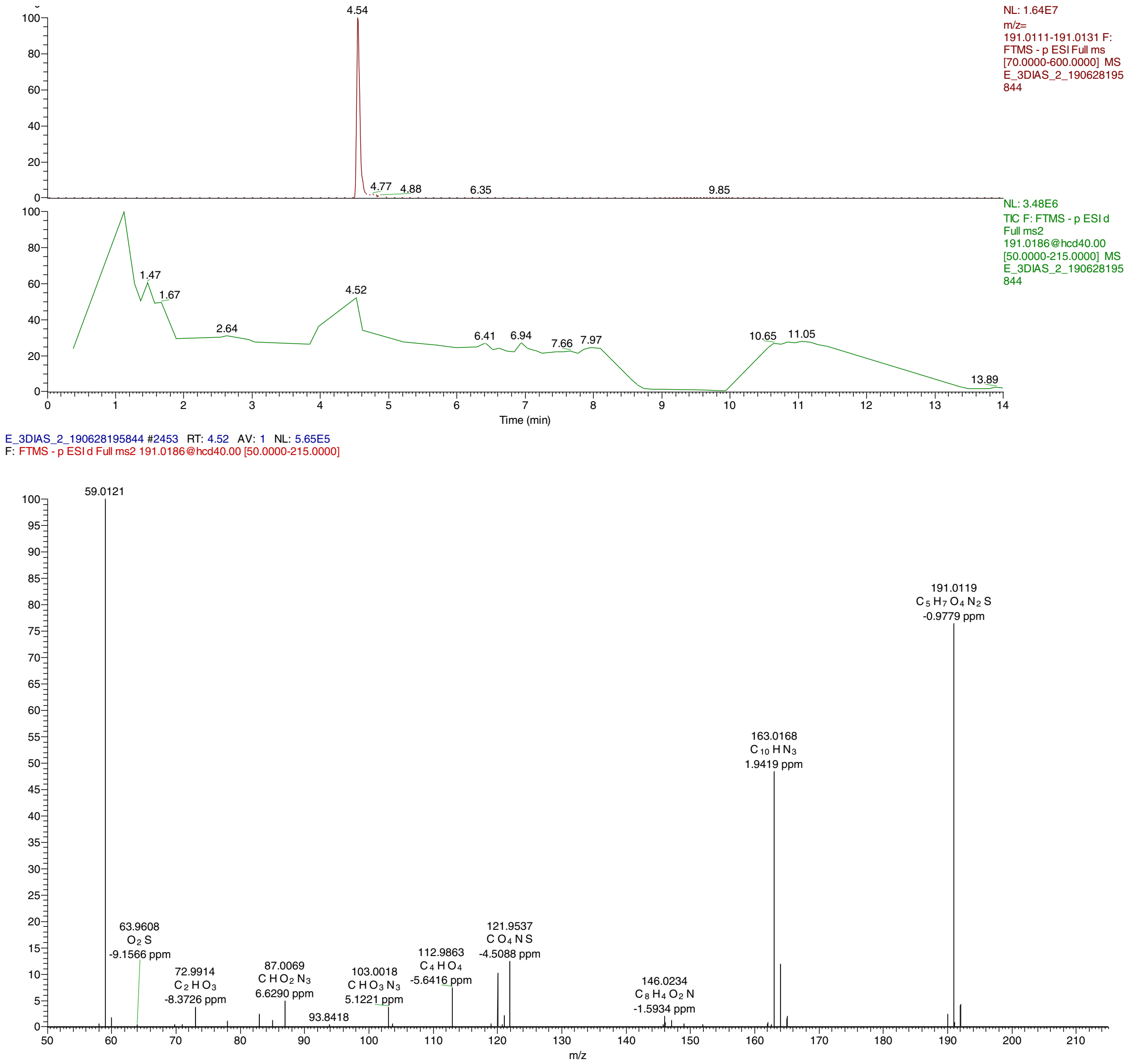

**Figure S18.** Extracted ion chromatograms (top: full MS, down: full MS$^2$) and MS$^2$ spectrum of *m/z* 191.0119 (TP192).

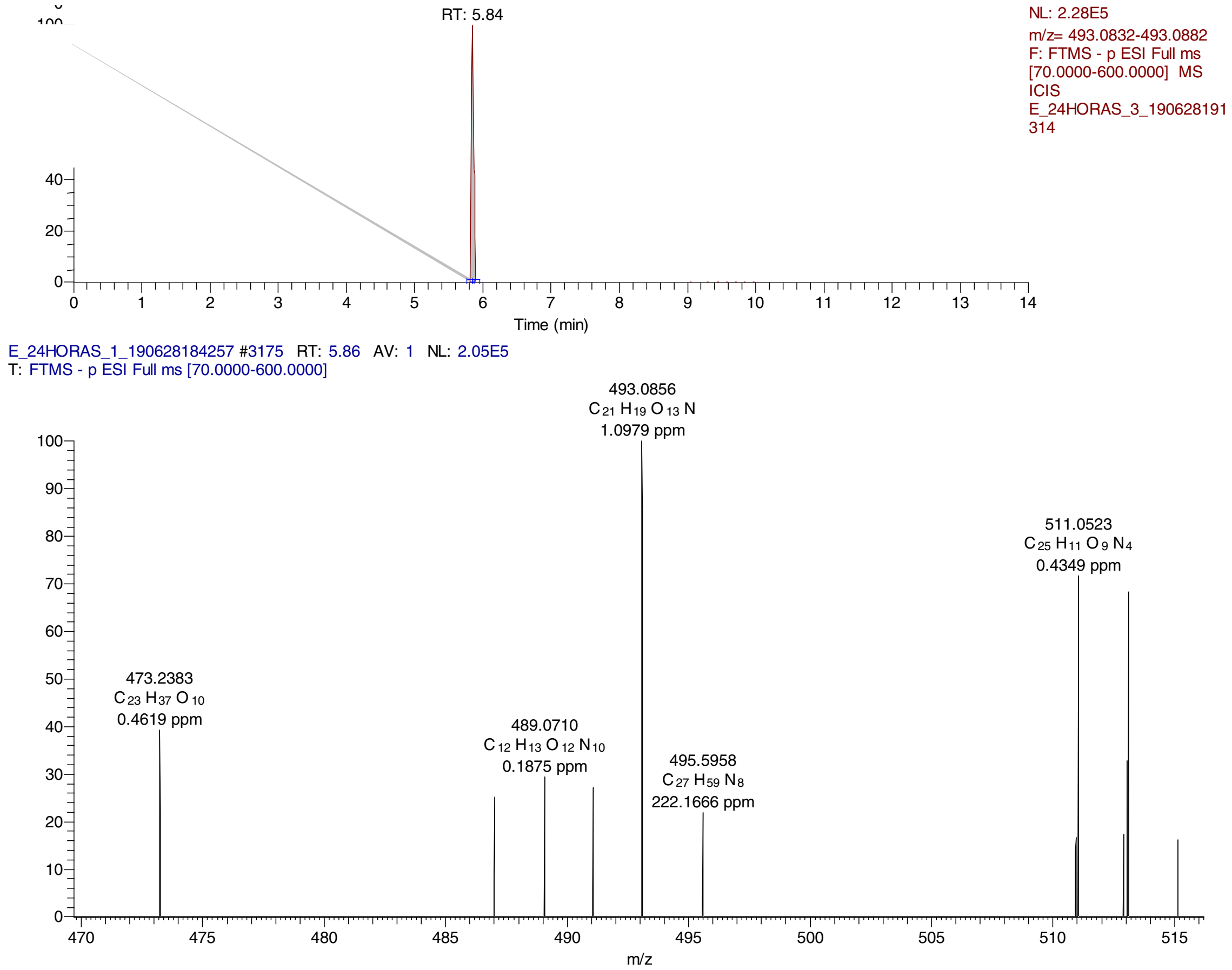

**Figure S19.** Extracted ion chromatogram and full MS spectrum of *m/z* 493.0856 (TP494).